\documentclass[sigconf,screen]{acmart}

\usepackage[normalem]{ulem}
\usepackage[bottom]{footmisc}

\usepackage{balance}

\usepackage{caption}
\usepackage{times}
\usepackage{oneinchmargins}
\usepackage{tightenum}

\usepackage{soul}

\usepackage[toc,page]{appendix}

\usepackage{subfig}

\usepackage{hyperref}
\hypersetup{hidelinks,colorlinks=true,linkcolor=red,citecolor=red}

\usepackage{setspace}
\usepackage{epsfig}
\usepackage{graphicx}
\usepackage{times}
\usepackage{amsmath}
\usepackage{url}
\usepackage{multirow}

\usepackage{tikz}
\usetikzlibrary{calc}

\usepackage{geometry}
\usepackage{xcolor}
\definecolor{commentgreen}{RGB}{2,112,10}
\definecolor{eminence}{RGB}{108,48,130}
\definecolor{weborange}{RGB}{255,165,0}
\definecolor{frenchplum}{RGB}{129,20,83}

\usepackage{listings}
\renewcommand{\em}{\it}

\newcommand{\x}{$\times$}

\newcommand{\ignore}[1]{}

\newcommand{\boldpara}[1]{\noindent{\textbf{#1}}}
\newcommand{\ulinebfpara}[1]{\noindent{\underline{\textbf{#1}}}}
\newcommand{\ulineitpara}[1]{\noindent{\underline{\textit{#1}}}}

\usepackage{array}
\newcolumntype{$}{>{\global\let\currentrowstyle\relax}}
\newcolumntype{^}{>{\currentrowstyle}}

\def\cfigure[#1,#2,#3]{
\begin{figure}
\vspace*{0mm}
\begin{center}

\includegraphics[width=3in]{#1} 
 
\vspace*{-3mm}\caption[]{#2
} \label{#3}
 
\vspace*{-5mm}
\end{center}
\end{figure}}

\def\cfigurefour[#1,#2,#3]{
\begin{figure}
\vspace*{0mm}
\begin{center}

\includegraphics[width=4in]{#1} 
 
\vspace*{-3mm}\caption[]{#2
} \label{#3}
 
\vspace*{-5mm}
\end{center}
\end{figure}}

\def\cfiguretemp[#1,#2,#3]{
\begin{figure}
\vspace*{0mm}
\begin{center}

\includegraphics[width=3.5in]{#1} 
 
\vspace*{-3mm}\caption[]{#2
} \label{#3}
 
\vspace*{-5mm}
\end{center}
\vspace*{-2mm}
\end{figure}}

\def\wfigure[#1,#2,#3]{
\begin{figure*}
\vspace*{0mm}
\begin{center}
 \includegraphics[width=\textwidth]{#1} 
 \vspace*{-3mm}\caption[]{#2
} \label{#3}
 
\end{center}
\end{figure*}}

\def\threefigure[#1,#2,#3,#4,#5]{
\begin{figure*}
\vspace*{0mm}
\begin{center}

\begin{tabular}{ccc}
\includegraphics[width=2in]{#1} & \includegraphics[width=2in]{#2} &  \includegraphics[width=2in]{#3} \\
(a) & (b) & (c) \\
\end{tabular}

\vspace*{-3mm}\caption[]{#4
} \label{#5}

\vspace*{-5mm}
\end{center}
\vspace*{-2mm}
\end{figure*}}

\def\dcfigure[#1,#2,#3,#4,#5,#6]{
{
\begin{figure*}
\begin{center}
\begin{minipage}[c]{\columnwidth}{
\includegraphics[width=\columnwidth]{#1} 
\vspace*{0mm}\caption[]{#2} \label{#3} \
}\end{minipage}\hspace*{\columnsep}\
\begin{minipage}[c]{\columnwidth}{
\includegraphics[width=\columnwidth]{#4} 
\vspace*{0mm}\caption[]{#5}\label{#6} \
}\end{minipage}
\end{center}
\end{figure*}
}
}

\def\tableByTable[#1,#2,#3,#4,#5,#6]{
{
\begin{table*}
\begin{center}
\begin{minipage}[c]{3in}{
\centering
{#1}
\vspace*{0mm}\tabcaption[]{#2}\label{#3} \
}\end{minipage}\hspace*{\columnsep}\
\begin{minipage}[c]{3in}{
\centering
{#4}
\vspace*{0mm}\tabcaption[]{#5}\label{#6} \
}\end{minipage}
\end{center}
\end{table*}
}
}

\def\figureByTable[#1,#2,#3,#4,#5,#6]{
{
\begin{figure*}
\begin{center}
\begin{minipage}[c]{3in}{
\centering
\includegraphics[width=\textwidth]{#1}
\vspace*{0mm}\figcaption[]{#2} \label{#3} \
}\end{minipage}\hspace*{\columnsep}\
\begin{minipage}[c]{3.3in}{
\centering
{#4}
\vspace*{0mm}\tabcaption[]{#5}\label{#6} \
}\end{minipage}
\end{center}
\end{figure*}
}
}

\def\tableByFigure[#1,#2,#3,#4,#5,#6]{
{
\begin{figure*}
\begin{center}
\begin{minipage}[c]{4.3in}{
\centering
{#1}
\vspace*{0mm}\tabcaption[]{#2} \label{#3} \
}\end{minipage}\hspace*{\columnsep}\
\begin{minipage}[c]{2.2in}{
\centering
\includegraphics[width=\textwidth]{#4}
\vspace*{-0.35in}\caption[]{#5}\label{#6} \
}\end{minipage}
\end{center}
\end{figure*}
}
}

\def\doublecfigure[#1,#2,#3,#4]{
{
\begin{figure}
\begin{center}
\begin{minipage}[c]{1.5in}{
\begin{center}
\includegraphics[width=1.5in]{#1}
\end{center}
}\end{minipage}\hspace*{1em}\
\begin{minipage}[c]{1.5in}{
\begin{center}
\includegraphics[width=1.5in]{#2}
\end{center}
}\end{minipage}
\vspace*{0mm}\caption[]{#3} \label{#4} \
\end{center}
\end{figure}
}
}

\def\qcfigure[#1,#2,#3,#4,#5,#6]{
{
\begin{figure*}
\vspace*{0.2in}\
\begin{center}
\begin{minipage}[c]{3in}{
\includegraphics[width=3in]{#1} 
\vspace*{-3mm}
}
\end{minipage}\hspace*{0.5in}\
\begin{minipage}[c]{3in}{
\includegraphics[width=3in]{#2} 
\vspace*{-3mm}
}\end{minipage}

\begin{minipage}[c]{3in}{
\includegraphics[width=3in]{#3} 
\vspace*{-3mm}
}
\end{minipage}\hspace*{0.5in}\
\begin{minipage}[c]{3in}{
\includegraphics[width=3in]{#4} 
\vspace*{-3mm}
}\end{minipage}
\end{center}
\caption[]{#5}\label{#6}
\end{figure*}
}
}

\def\twfigure[#1,#2,#3,#4,#5]{
{
\begin{figure*}
\vspace*{0.2in}\
\begin{center}
\begin{minipage}[c]{6.5in}{
\includegraphics[width=6.5in]{#1} 
\vspace*{-3mm}
}
\end{minipage}

\begin{minipage}[c]{6.5in}{
\includegraphics[width=6.5in]{#2} 
\vspace*{-3mm}
}\end{minipage}

\begin{minipage}[c]{6.5in}{
\includegraphics[width=6.5in]{#3} 
\vspace*{-3mm}
}
\end{minipage}
\end{center}
\caption[]{#4}\label{#5}
\end{figure*}
}
}

\def\dwfigure[#1,#2,#3,#4]{
{
\begin{figure*}
\vspace*{0.2in}\
\begin{center}
\begin{minipage}[c]{6.5in}{
\includegraphics[width=6.5in]{#1} 
\vspace*{-3mm}
}
\end{minipage}

\begin{minipage}[c]{6.5in}{
\includegraphics[width=6.5in]{#2} 
\vspace*{-3mm}
}\end{minipage}

\end{center}
\caption[]{#3}\label{#4}
\end{figure*}
}
}

\def\dssfigure[#1,#2,#3,#4,#5,#6]{
{
\begin{figure*}
\vspace*{0.2in}\
\begin{center}
\begin{minipage}[c]{4in}{
\includegraphics[width=4in]{#1}
\vspace*{-3mm}\caption[]{#2} \label{#3} \
}\end{minipage}\hspace*{0.5in}\
\begin{minipage}[c]{2in}{
\includegraphics[width=2in]{#4}
\vspace*{-3mm}\caption[]{#5}\label{#6} \
}\end{minipage}
\end{center}
\vspace*{-0.4in}\
\end{figure*}
}
}

\def\dsfigure[#1,#2,#3,#4,#5,#6]{
{
\begin{figure*}
\vspace*{0.2in}\
\begin{center}
\begin{minipage}[c]{3in}{
\includegraphics[width=3in]{#1}
\vspace*{-3mm}\caption[]{#2} \label{#3} \
}\end{minipage}\hspace*{0.5in}\
\begin{minipage}[c]{3in}{
\hspace*{0.5in}\
\includegraphics[height=3in]{#4}
\vspace*{-3mm}\caption[]{#5}\label{#6} \
}\end{minipage}
\end{center}
\vspace*{-0.4in}\
\end{figure*}
}
}

\def\dsyfigure[#1,#2,#3,#4,#5,#6]{
{
\begin{figure*}
\vspace*{0.2in}\
\begin{center}
\begin{minipage}[c]{2.5in}{
\includegraphics[height=2.5in]{#1}
\vspace*{-3mm}\caption[]{#2} \label{#3} \
}\end{minipage}\hspace*{0.5in}\
\begin{minipage}[c]{2.5in}{
\includegraphics[height=2.5in]{#4}
\vspace*{-3mm}\caption[]{#5}\label{#6} \
}\end{minipage}
\end{center}
\vspace*{-0.4in}\
\end{figure*}
}
}

\def\dyfigure[#1,#2,#3,#4,#5,#6]{
{
\begin{figure*}
\vspace*{0.2in}\
\begin{center}
\begin{minipage}[c]{3in}{
\includegraphics[height=3in]{#1} 
\vspace*{-3mm}\caption[]{#2} \label{#3} \
}\end{minipage}\hspace*{0.5in}\
\begin{minipage}[c]{3in}{
\includegraphics[height=3in]{#4} 
\vspace*{-3mm}\caption[]{#5}\label{#6} \
}\end{minipage}
\end{center}
\vspace*{-0.4in}\
\end{figure*}
}
}

\def\dyoldfigure[#1,#2,#3,#4,#5,#6]{
{
\begin{figure*}
\vspace*{0.2in}\
\begin{center}
\begin{minipage}[c]{3in}{
\epsfysize=2.0in\
\hspace{0.5in}\
\epsfbox{#1}
\vspace*{-3mm}\caption[]{#2} \label{#3} \
}\end{minipage}\hspace*{0.25in}\
\begin{minipage}[c]{3in}{
\epsfysize=2.0in\
\hspace{0.5in}\
\epsfbox{#4}
\vspace*{-3mm}\caption[]{#5}\label{#6} \
}\end{minipage}
\end{center}
\vspace*{-0.4in}\
\end{figure*}
}
}

\def\cfiguredouble[#1,#2,#3,#4]{
\begin{figure}
\vspace*{0.2in}\
\begin{center}
\begin{minipage}[c]{1.5in}{
\epsfxsize=1.5in\
\epsfbox{#1}
}\end{minipage}\hspace*{0.1in}\
\begin{minipage}[c]{1.5in}{
\epsfxsize=1.5in\
\vspace{0.1in}\epsfbox{#2}
}\end{minipage}\vspace*{-0.10in} \caption[]{#3}\label{#4}
\end{center}
\vspace*{-0.4in}\
\end{figure}
}

\def\wpfigure[#1,#2,#3,#4]{
\begin{figure*}
\vspace*{4mm}
\begin{center}

\includegraphics[width=#4]{#1} 

\vspace*{-3mm}\caption[]{#2
} \label{#3}

\vspace*{-5mm}
\end{center}
\end{figure*}}

\def\wprfigure[#1,#2,#3,#4,#5]{
\begin{figure*}
\vspace*{4mm}
\begin{center}

\includegraphics[width=#4, angle=#5]{#1} 

\vspace*{-3mm}\caption[]{#2
} \label{#3}

\vspace*{-5mm}
\end{center}
\end{figure*}}

\def\DoubleFigureWSlide[#1,#2,#3,#4,#5,#6,#7,#8,#9]{
\begin{figure*}
\vspace*{#9}
\begin{center}
\begin{minipage}{#4}
\includegraphics[width=#4]{#1}
\vspace*{-3mm}\caption{#2
}\label{#3}
\end{minipage}
\hspace{2em}
\begin{minipage}{#8}
\includegraphics[width=#8]{#5}
\vspace*{-3mm}\caption{#6
}\label{#7}
\end{minipage}
\vspace*{-5mm}
\end{center}
\end{figure*}
}

\def\DoubleFigureW[#1,#2,#3,#4,#5,#6,#7,#8]{
\begin{figure*}
\vspace*{0in}
\begin{center}
\begin{minipage}{#4}
\includegraphics[width=#4]{#1}
\vspace*{-3mm}\caption{#2
}\label{#3}
\end{minipage}
\hspace{2em}
\begin{minipage}{#8}
\includegraphics[width=#8]{#5}
\vspace*{-3mm}\caption{#6
}\label{#7}
\end{minipage}
\vspace*{-5mm}
\end{center}
\end{figure*}
}

\def\DoubleFigureWHack[#1,#2,#3,#4,#5,#6,#7,#8]{
\begin{figure*}
\vspace*{0in}
\begin{center}
\begin{minipage}{3in}
\includegraphics[width=#4]{#1}
\vspace*{-3mm}\caption{#2
}\label{#3}
\end{minipage}
\hspace{2em}
\begin{minipage}{3in}
\includegraphics[width=#8]{#5}
\vspace*{-3mm}\caption{#6
}\label{#7}
\end{minipage}
\vspace*{-5mm}
\end{center}
\end{figure*}
}

\def\ddcfigure[#1,#2,#3,#4]{
\begin{figure*}
\vspace*{0.2in}\
\begin{center}
\begin{minipage}[c]{\columnwidth}{
\includegraphics[width=\columnwidth]{#1} 
}\end{minipage}\hspace{0.5in}\
\begin{minipage}[c]{\columnwidth}{
\includegraphics[width=\columnwidth]{#2} 
}\end{minipage} \caption[]{#3}\label{#4}
\end{center}
\end{figure*}
}

\def\ddcfigureSlide[#1,#2,#3,#4,#5]{
\begin{figure*}
\vspace*{#5}\
\begin{center}
\begin{minipage}[c]{3in}{
\includegraphics[height=3in]{#1} 
}\end{minipage}\hspace{0.5in}\
\begin{minipage}[c]{3in}{
\includegraphics[height=3in]{#2} 
}\end{minipage}\vspace*{-0.10in} \caption[]{#3}\label{#4}
\end{center}
\vspace*{-0.4in}\
\end{figure*}
}

\def\cxfigure[#1,#2,#3]{
\begin{figure}
\vspace*{4mm}
\begin{center}
 
\epsfxsize=2.5in\
\epsfbox{#1}\
 
\vspace*{-0.10in}\caption[]{#2
} \label{#3}
 
\vspace*{-5mm}
\end{center}
\vspace*{-2mm}
\end{figure}}

\newcommand{\figWidthSix}{1.72in}

\newcommand{\beforecaption}{\vspace{-.15cm}\begin{spacing}{0.85}}
\newcommand{\aftercaption}{\vspace{-.45cm}\end{spacing}}

\newcommand{\mycaption}[3]{\beforecaption\caption{\label{#1}{\bf \small #2} \em\footnotesize #3}\aftercaption}

\newcommand{\eg}{\textit{e.g.}}
\newcommand{\ie}{\textit{i.e.}}

\newcommand{\KB}{\,KB}
\newcommand{\MB}{\,MB}
\newcommand{\GB}{\,GB}
\newcommand{\Gbps}{\,Gbps}
\newcommand{\TB}{\,TB}

\newcommand{\mus}{\mbox{\,$\mu s$}}
\newcommand{\ms}{\mbox{\,$ms$}}
\newcommand{\ns}{\mbox{\,$ns$}}

\newcounter{reqs}
\newcounter{case}
\newcommand{\sys}{Clio}
\newcommand{\syskv}{Clio-KV}
\newcommand{\sysmv}{Clio-MV}
\newcommand{\sysdf}{Clio-DF}
\newcommand{\pdm}{PDM}

\newcommand{\sysboard}{CBoard}
\newcommand{\syslib}{CLib}
\newcommand{\cas}{{\texttt{rCAS}}}
\newcommand{\tas}{{\texttt{rTAS}}}

\newcommand{\alloc}{\texttt{ralloc}}
\newcommand{\sysfree}{\texttt{rfree}}
\newcommand{\sysread}{\texttt{rread}}
\newcommand{\syswrite}{\texttt{rwrite}}
\newcommand{\syslock}{\texttt{rlock}}
\newcommand{\sysunlock}{\texttt{runlock}}

\newcommand{\fence}{\texttt{rfence}}
\newcommand{\release}{\texttt{rrelease}}
\newcommand{\pid}{CPID}
\newcommand{\CN}{CN}
\newcommand{\MN}{MN}
\newcommand{\md}{MemDisagg}

\newcommand{\rspace}{RAS}
\newcommand{\poll}{\texttt{rpoll}}

\newif\ifremark
\long\def\remark#1{
\ifremark%
        \begingroup%
        \dimen0=\columnwidth
        \advance\dimen0 by -1in%
        \setbox0=\hbox{\parbox[b]{\dimen0}{\protect\em #1}}
        \dimen1=\ht0\advance\dimen1 by 2pt%
        \dimen2=\dp0\advance\dimen2 by 2pt%
        \vskip 0.25pt%
        \hbox to \columnwidth{%
                \vrule height\dimen1 width 3pt depth\dimen2%
                \hss\copy0\hss%
                \vrule height\dimen1 width 3pt depth\dimen2%
        }%
        \endgroup%
\fi}

\remarktrue

\setcopyright{rightsretained}
\acmPrice{}
\acmDOI{10.1145/3503222.3507762}
\acmYear{2022}
\copyrightyear{2022}
\acmSubmissionID{asplos22main-p968-p}
\acmISBN{978-1-4503-9205-1/22/02}
\acmConference[ASPLOS '22]{Proceedings of the 27th ACM International Conference on Architectural Support for Programming Languages and Operating Systems}{February 28 -- March 4, 2022}{Lausanne, Switzerland}
\acmBooktitle{Proceedings of the 27th ACM International Conference on Architectural Support for Programming Languages and Operating Systems (ASPLOS '22), February 28 -- March 4, 2022, Lausanne, Switzerland}


\begin{document}

\pagenumbering{arabic}
\pagestyle{plain}

\definecolor{pink}{rgb}{1.0,0.47,0.6}

\newcommand{\fixme}[1]{{\color{red}\textbf{#1}}}

\newcommand{\yiying}[1]{{\color{blue}\textbf{#1}}}
\newcommand{\zhiyuan}[1]  {\noindent{\color{green} {\bf \fbox{Zhiyuan}     {\it#1}}}}
\newcommand{\yutong}[1]{{\color{pink}\textbf{#1}}}
\newcommand{\xuhao}[1]{{\color{yellow}\textbf{#1}}}
\newcommand{\yizhou}[1]  {\noindent{\color{blue} {\bf \fbox{YS}     {\it#1}}}}

\newcommand{\DEL}[1]  {\noindent{\color{red} {\bf \fbox{DEL}     {\it \sout{ #1 }}}}}

\newcommand{\note}[2]{\fixme{$\ll$ #1 $\gg$ #2}}

\newcommand{\shinyeh}[1]{{\color{purple}\textbf{#1}}}

\title{\sys: A Hardware-Software Co-Designed Disaggregated Memory System}

\author{Zhiyuan Guo}
\authornote{Both authors contributed equally to the paper}
\affiliation{%
 \institution{University of California, San Diego}
 \city{San Diego}
 \state{California}
 \country{USA}}
\email{z9guo@ucsd.edu}

\author{Yizhou Shan}
\authornotemark[1]
\affiliation{%
 \institution{University of California, San Diego}
 \city{San Diego}
 \state{California}
 \country{USA}}
\email{ys@ucsd.edu}

\author{Xuhao Luo}
\affiliation{%
 \institution{University of California, San Diego}
 \city{San Diego}
 \state{California}
 \country{USA}}
\email{x3luo@ucsd.edu}

\author{Yutong Huang}
\affiliation{%
 \institution{University of California, San Diego}
 \city{San Diego}
 \state{California}
 \country{USA}}
\email{yutonghuang@ucsd.edu}

\author{Yiying Zhang}
\affiliation{%
 \institution{University of California, San Diego}
 \city{San Diego}
 \state{California}
 \country{USA}}
\email{yiying@ucsd.edu}


\begin{abstract}

Memory disaggregation has attracted great attention recently because of its benefits in efficient memory utilization and ease of management. So far, memory disaggregation research has all taken one of two approaches: building/emulating memory nodes using regular servers or building them using raw memory devices with no processing power. The former incurs higher monetary cost and faces tail latency and scalability limitations, while the latter introduces performance, security, and management problems.

Server-based memory nodes and memory nodes with no processing power are two extreme approaches. We seek a sweet spot in the middle by proposing a hardware-based memory disaggregation solution that has the right amount of processing power at memory nodes. Furthermore, we take a clean-slate approach by starting from the requirements of memory disaggregation and designing a \textit{memory-disaggregation-native} system.

We built \textit{Clio}, a disaggregated memory system that virtualizes, protects, and manages disaggregated memory at hardware-based memory nodes. The Clio hardware includes a new virtual memory system, a customized network system, and a framework for computation offloading. In building Clio, we not only co-design OS functionalities, hardware architecture, and the network system, but also co-design compute nodes and memory nodes. Our FPGA prototype of Clio demonstrates that each memory node can achieve 100\,Gbps throughput and an end-to-end latency of 2.5\mbox{\,$\mu s$} at median and 3.2\mbox{\,$\mu s$} at the 99th percentile. Clio also scales much better and has orders of magnitude lower tail latency than RDMA. It has 1.1$\times$ to 3.4$\times$ energy saving compared to CPU-based and SmartNIC-based disaggregated memory systems and is 2.7$\times$ faster than software-based SmartNIC solutions.

\end{abstract}
\maketitle

\section{Introduction}
\label{sec:introduction}

Modern datacenter applications like graph computing, data analytics, and deep learning have an increasing demand for access to large amounts of memory~\cite{FastSwap}.
Unfortunately, servers are facing {\em memory capacity walls} because of pin, space, and power limitations~\cite{HP-MemoryEvol,ITRS14,MemoryWall95}.
Going forward, it is imperative for datacenters to seek solutions that can go beyond what a (local) machine can offer, \ie, using remote memory.
At the same time, datacenters are seeing the needs from management and resource utilization perspectives
to {\em disaggregate} resources~\cite{Ali-SinglesDay,SnowFlake-NSDI20,FB1}\textemdash separating hardware resources into different network-attached pools 
that can be scaled and managed independently.
These real needs have pushed the idea of memory disaggregation ({\em \md} for short):
organizing computation and memory resources as two separate network-attached
pools, one with compute nodes ({\em CN}s) and one with memory nodes ({\em MN}s).

So far, \md\ researches have all taken one of two approaches: 
building/emulating \MN{}s using regular servers~\cite{AIFM,InfiniSwap,FastSwap,Shan18-OSDI,zombieland} 
or using raw memory devices with no processing power~\cite{Tsai20-ATC,Lim09-disaggregate,Lim12-HPCA,HP-TheMachine}.
The fundamental issues of server-based approaches such as RDMA-based systems are the monetary and energy cost of a host server and the inherent performance and scalability limitations caused by the way NICs interact with the host server's virtual memory system.
Raw-device-based solutions have low costs.
However, they introduce performance, security, and management problems
because when \MN{}s have no processing power, all the data and control planes have to be handled at \CN{}s~\cite{Tsai20-ATC}.

Server-based \MN{}s and \MN{}s with no processing power are two extreme approaches of building \MN{}s.
We seek a sweet spot in the middle by proposing a hardware-based \md\ solution that has the right amount of processing power at \MN{}s.
Furthermore, we take a clean-slate approach by starting from the requirements of \md\
and designing a \md-native system.

We built {\em \sys}\footnote{Clio is the daughter of Mnemosyne, the Greek goddess of memory.}, a hardware-based disaggregated memory system.
\sys\ includes a \CN-side user-space library called {\em \syslib}
and a new hardware-based \MN\ device called {\em \sysboard}.
Multiple application processes running on different \CN{}s can allocate memory from the same \sysboard, with each process having its own {\em remote virtual memory address space}.
Furthermore, one remote virtual memory address space can span multiple \sysboard{}s.
Applications can perform byte-granularity remote memory read/write and use \sys's synchronization primitives for synchronizing concurrent accesses to shared remote memory .

A key research question in designing \sys\ is \textit{\textbf{how to use limited hardware resources to achieve 100\Gbps, microsecond-level average and tail latency for TBs of memory and thousands of concurrent clients?}}
These goals are important and unique for \md.
A good \md\ solution should reduce the total CapEx and OpEx costs compared to traditional non-disaggregated systems and thus cannot afford to use large amounts of hardware resources at \MN{}s.
Meanwhile, remote memory accesses should have high throughput and low average and tail latency, because even after caching data at \CN-local memory, there can still be fairly frequent accesses to \MN{}s and the overall application performance can be impacted if they are slow~\cite{disagg-osdi16}.
Finally, unlike traditional single-server memory, a disaggregated \MN\ should allow many \CN{}s to store large amounts of data so that we only need a few of them to reduce costs and connection points in a cluster.
How to achieve each of the above cost, performance, and scalability goals {\em individually} is relatively well understood.
However, achieving all these seemingly conflicting goals {\em simultaneously} is hard and previously unexplored.

Our main idea is to \textbf{\textit{eliminate state from the \MN\ hardware}}.
Here, we overload the term ``state elimination'' with two meanings: 1) the \MN\ can treat each of its incoming requests in isolation even if requests that the client issues can sometimes be inter-dependent, 
and 2) the \MN\ hardware does not store metadata or deals with it.
Without remembering previous requests or storing metadata, an \MN\ would only need a tiny amount of on-chip memory that does not grow with more clients, thereby {\em saving monetary and energy cost} and achieving {\em great scalability}.
Moreover, without state, the hardware pipeline can be made {\em smooth} and {\em performance deterministic}.
A smooth pipeline means that the pipeline does not stall, which is only possible if requests do not need to wait for each other.
It can then take one incoming data unit from the network every fixed number of cycles (1 cycle in our implementation), achieving constantly {\em high throughput}.
A performance-deterministic pipeline means that the hardware processing does not need to wait for any slower metadata operations and thus has {\em bounded tail latency}.

Effective as it is, can we really eliminate state from \MN\ hardware? 
First, as with any memory systems, users of a disaggregate memory system expect it to deliver certain reliability and consistency guarantees (\eg, a successful write should have all its data written to remote memory, a read should not see the intermediate state of a write, etc.). 
Implementing these guarantees requires proper ordering among requests and involves state even on a single server. 
The network separation of disaggregated memory would only make matters more complicated.
Second, quite a few memory operations involve metadata, and they too need to be supported by disaggregated memory.
Finally, many memory and network functionalities are traditionally associated with a client process and involve per-process/client metadata (\eg, one page table per process, one connection per client, etc.). 
Overcoming these challenges require the re-design of traditional memory and network systems.

Our first approach is to separate the metadata/control plane and the data plane, with the former running as software on a low-power ARM-based SoC at \MN\ and the latter in hardware at \MN. 
Metadata operations like memory allocation usually need more memory but are rarer (thus not as performance critical) compared to data operations.
A low-power SoC's computation speed and its local DRAM are sufficient for metadata operations.
On the other hand, data operations (\ie, all memory accesses) should be fast and are best handled purely in hardware. 
Even though the separation of data and control plane is a common technique that has been applied in many areas~\cite{4d-sdn,netvirt,arrakis}, a separation of memory system control and data planes has not been explored before and is not easy, as we will show in this paper.

Our second approach is to re-design the memory and networking data plane so that most state can be managed only at the \CN\ side.
Our observation here is that the \MN{} only {\em responds} to memory requests but never {\em initiates} any.
This \CN-request-\MN-respond model allows us to use a custom, connection-less reliable transport protocol that implements almost all transport-layer services and state at \CN{}s, allowing \MN{}s to be free from traditional transport-layer processing.
Specifically, our transport protocol manages request IDs, transport logic, retransmission buffer, congestion, and incast control all at \CN{}s. It provides 
reliability by ordering and retrying an entire memory request at the \CN\ side.
As a result, the \MN{} does not need to worry about per-request state or inter-request ordering and only needs a tiny amount of hardware resources which do not grow with the number of clients.
 
With the above two approaches, the hardware can be largely simplified and thus cheaper, faster, and more scalable.
However, we found that \textit{\textbf{complete state elimination at \MN{}s is neither feasible nor ideal}}. To ensure correctness, the \MN\ has to maintain some state (\eg, to deal with non-idempotent operations). To ensure good data-plane performance, not every operation that involves state should be moved to the low-power SoC or to \CN{}s.
Thus, our approach is to eliminate as much state as we can without affecting performance or correctness and to carefully design the remaining state so that it causes small and bounded space and performance overhead.

For example, we perform paging-based virtual-to-physical memory address mapping and access permission checking at the \MN\ hardware pipeline, as these operations are needed for every data access.
Page table is a kind of state that could potentially cause performance and scalability issues but has to be accessed in the data path.
We propose a new overflow-free, hash-based page table design where 1) all page table lookups have bounded and low latency (at most one DRAM access time in our implementation), and 2) the total size of all page table entries does not grow with the number of client processes.
As a result, even though we cannot eliminate page table from the \MN\ hardware, we can still meet our cost, performance, or scalability requirements.

Another data-plane operation that involves metadata is page fault handling, which is a relatively common operation because we allocate physical memory on demand.
Today's page fault handling process is slow and involves metadata for physical memory allocation.
We propose a new mechanism to handle page faults in hardware and finish all the handling within bounded hardware cycles.
We make page fault handling performance deterministic by moving  physical memory allocation operations to software running at the SoC.
We further move these allocation operations off the performance-critical path by pre-generating free physical pages to a fix-sized buffer that the hardware pipeline can pull when handling page faults.

We prototyped \sysboard\ with a small set of Xilinx ZCU106 MPSoC FPGA boards~\cite{ZCU106} and built three applications using \sys:
a FaaS-style image compression utility, a radix-tree index, and a key-value store.
We compared \sys\ with native RDMA, two RDMA-based disaggregated/remote memory systems~\cite{Tsai20-ATC,Kalia14-RDMAKV}, 
a software emulation of hardware-based disaggregated memory~\cite{Shan18-OSDI},
and a software-based SmartNIC~\cite{BlueField}.
\sys\ scales much better and has orders of magnitude lower tail latency than RDMA, 
while achieving similar throughput and median latency as RDMA (even with the slower FPGA frequency in our prototype).
\sys\ has 1.1\x\ to 3.4\x\ energy saving compared to CPU-based and SmartNIC-based disaggregated memory systems 
and is 2.7\x\ faster than SmartNIC solutions. 
\sys\ is publicly available at \url{https://github.com/WukLab/Clio}.

\section{Goals and Related Works}
\label{sec:motivation}

Resource disaggregation 
separates different types of resources into different pools,
each of which can be independently managed and scaled.
Applications can allocate resources from any node in a resource pool, resulting in tight resource packing. 
Because of these benefits, 
many datacenters have adopted the idea of disaggregation, often at the storage 
layer~\cite{FACEBOOK-BRYCECANYON,FB1,SnowFlake-NSDI20,AMAZON-S3,AMAZON-EBS,Pangu,FC-SAN-book}.
With the success of disaggregated storage, 
researchers in academia and industry have also sought ways to disaggregate memory
(and persistent memory)
\cite{Lim09-disaggregate,FireBox-FASTKeynote,IntelRackScale,Lim12-HPCA,Shan18-OSDI,Shan17-SOCC,RAMCloud,Tsai20-ATC,AIFM,FastSwap,InfiniSwap,Semeru,Nitu18-EUROSYS}.
Different from storage disaggregation,
\md\ needs to achieve at least an order of magnitude higher performance and it should offer a byte-addressable interface.
Thus, \md\ poses new challenges and requires new designs.
This section discusses the requirements of \md\ and why existing solutions cannot fully meet them.

\subsection{\md\ Design Goals}
\label{sec:requirements}
In general, \md\ has the following features, some of which are hard requirements while others are desired goals.

\stepcounter{reqs}
\boldpara{R\arabic{reqs}: Hosting large amounts of memory with high utilization.}
To keep the number of memory devices and total cost of a cluster low,
each \MN\ should host hundreds GBs to a few TBs of memory that is expected to be close to fully utilized.
To most efficiently use the disaggregated memory, we should allow applications to create and access {\em disjoint} memory regions of arbitrary sizes at \MN.

\stepcounter{reqs}
\boldpara{R\arabic{reqs}: Supporting a huge number of concurrent clients.}
To ensure tight and efficient resource packing,
we should allow many (\eg, thousands of) client processes running on tens of \CN{}s to access and share an \MN.
This scenario is especially important for new data-center trends like serverless computing and microservices where applications run as large amounts of small units.

\stepcounter{reqs}
\boldpara{R\arabic{reqs}: Low-latency and high-throughput.}
We envision future systems to have a new memory hierarchy, where disaggregated memory is larger and slower than local memory but still faster than storage.
Since \md\ is network-based, a reasonable performance target of it is to match the state-of-the-art network speed,
\ie, 100\Gbps\ throughput (for bigger requests) and sub-2\mus\ median end-to-end latency (for smaller requests).
 
\stepcounter{reqs}
\boldpara{R\arabic{reqs}: Low tail latency.}
Maintaining a low tail latency is important in meeting service-level objectives (SLOs) in data centers.
Long tails like RDMA's 16.8\ms\ remote memory access can be detrimental to applications that are short running (\eg, serverless computing workloads) or have large fan-outs or big DAGs
(because they need to wait for the slowest step to finish)~\cite{taillatency}.

\stepcounter{reqs}
\boldpara{R\arabic{reqs}: Protected memory accesses.}
As an \MN{} can be shared by multi-tenant applications running at \CN{}s, 
we should properly isolate memory spaces used by them.
Moreover, to prevent buggy or malicious clients from reading/writing arbitrary memory at \MN{}s, we should not allow the direct access of \MN{}s' physical memory from the network and \MN{}s should check the access permission.

\stepcounter{reqs}
\boldpara{R\arabic{reqs}: Low cost.}
A major goal and benefit of resource disaggregation is cost reduction.
A good \md\ system should have low {\em overall} CapEx and OpEx costs.
Such a system thus should not 1) use expensive hardware to build \MN{}s, 
2) consume huge energy at \MN{}s,
and 3) add more costs at \CN{}s than the costs saved at \MN{}s.

\stepcounter{reqs}
\boldpara{R\arabic{reqs}: Flexible.}
With the fast development of datacenter applications, hardware, and network, a sustainable \md\ solution should be flexible and extendable,
for example, to support high-level APIs like pointer chasing~\cite{AIFM,Aguilera-FarMemory},
to offload some application logic to memory devices~\cite{AIFM,StRoM},
or to incorporate different network transports~\cite{Homa,NDP,TONIC} and congestion control algorithms~\cite{swift-sigcomm,1RMA,hpcc-sigcomm19}.

%


\subsection{Server-Based Disaggregated Memory}
\label{sec:rdma}

\md\ research so far has mainly taken a server-based approach by using regular servers as \MN{}s~\cite{InfiniSwap,FastSwap,Semeru,Shan18-OSDI,AIFM,zombieland,FaRM},
usually on top of RDMA.
The common limitation of these systems is their reliance on a host server and the resulting CPU energy costs, both of which violate \textbf{R6}.

\ulinebfpara{RDMA} is what most server-based \md\ solutions are based on, with some using RDMA for swapping memory between \CN{}s and \MN{}s~\cite{InfiniSwap,FastSwap,Semeru} and some using RDMA for explicitly accessing \MN{}s~\cite{AIFM,zombieland,FaRM}.
Although RDMA has low average latency and high throughput, it has a set of scalability and tail-latency problems.

A process ($P_M$) running at an \MN\ needs to allocate memory in its virtual memory address space 
and {\em register} the allocated memory (called a memory region, or MR) with the RDMA NIC (RNIC).
The host OS and MMU set up and manage the page table that maps $P_M$'s virtual addresses ({\em VA}s) to physical memory addresses ({\em PA}s).
To avoid always accessing host memory for address mapping, RNICs cache page table entries (PTEs),
but when more PTEs are accessed than what this cache can hold, RDMA performance degrades significantly (Figure~\ref{fig-pte-mr} and \cite{FaRM,Tsai17-SOSP}).
Similarly, RNICs cache MR metadata and incur degraded performance when the cache is full. 
Thus, RDMA has serious performance issues with either large memory (PTEs) or many disjoint memory regions (MRs), violating \textbf{R1}.
Moreover, RDMA uses a slow way to support on-demand allocation: the RNIC interrupts the host OS for handling page faults.
From our experiments, a faulting RDMA access is 14100\x\ slower than a no-fault access (violating \textbf{R4}).

To mitigate the above performance and scalability issues, most RDMA-based systems today~\cite{FaRM,Tsai17-SOSP} 
preallocate a big MR with huge pages and pin it in physical memory.
This results in inefficient memory space utilization and violates \textbf{R1}.
Even with this approach, there can still be a scalability issue (\textbf{R2}),
as RDMA needs to create at least one MR for each protection domain (\ie, each client).

In addition to problems caused by RDMA's memory system design, reliable RDMA, the mode used by most \md\ solutions, suffers from a connection queue pair (QP) scalability issue, also violating \textbf{R2}.
Finally, today's RNICs violate \textbf{R7} because of their rigid one-sided RDMA interface and the close-sourced, hardware-based transport implementation.
Solutions like 1RMA~\cite{1RMA} and IRN~\cite{IRN} mitigate the above issues by either onloading part of the transport back to software or proposing a new hardware design.

\ulinebfpara{LegoOS}~\cite{Shan18-OSDI}, our own previous work, is a distributed operating system designed for resource disaggregation.
Its \MN{} includes a virtual memory system that maps VAs of application processes running at \CN{}s to \MN\ PAs. \sys's \MN{} performs the same type of address translation.
However, LegoOS emulates \MN\ devices using regular servers and we built its virtual memory system in software,
which has a stark difference from a hardware-based virtual memory system. 
For example, LegoOS uses a thread pool that handles incoming memory requests by looking up a hash table for address translation and permission checking.
This software approach is the major performance bottleneck in LegoOS (\S\ref{sec:results}),
violating \textbf{R3}.
Moreover, LegoOS 
uses RDMA for its network communication hence inheriting its limitations.

\subsection{Physical Disaggregated Memory}
\label{sec:pdm}

One way to build \md\ without a host server is to treat it as raw, physical memory,
a model we call {\em \pdm}.
The \pdm\ model has been adopted by a set of coherent interconnect proposals~\cite{Genz-citation,CXL-citation},
HPE's Memory-Driven Computing project~\cite{HP-TheMachine,THEMACHINE-HOTOS,HP-MODC-POSTER,THEMACHINE-WEB}.
A recent disaggregated hashing system~\cite{race-atc21} and our own recent work on disaggregated key-value systems~\cite{Tsai20-ATC} also adopt the \pdm\ model and emulate remote memory with regular servers.
To prevent applications from accessing raw physical memory,
these solutions add an indirection layer at \CN{}s in hardware~\cite{Genz-citation,CXL-citation} or software~\cite{Tsai20-ATC,race-atc21}
to map client process VAs or keys
to \MN\ PAs. 

There are several common problems with all the \pdm\ solutions.
First, because \MN{}s in \pdm\ are raw memory, \CN{}s need multiple network round trips to access an \MN\ 
for complex operations like pointer chasing and concurrent operations that need synchronization~\cite{Tsai20-ATC}, violating \textbf{R3} and \textbf{R7}.
Second, \pdm\ requires the client side to manage disaggregated memory.
For example, \CN{}s need to coordinate with each other or use a global server~\cite{Tsai20-ATC} to perform tasks like memory allocation.
Non-\MN-side processing is much harder, performs worse compared to memory-side management (violating \textbf{R3}), and could even result in higher overall costs because of the high computation added at \CN{}s (violating \textbf{R6}).
Third, exposing physical memory makes it hard to provide security guarantees (\textbf{R5}),
as \MN{}s have to authenticate that every access is to a legit physical memory address belonging to the application.
Finally, all existing \pdm\ solutions require physical memory pinning at \MN{}s, causing memory wastes and violating \textbf{R1}.

In addition to the above problems, none of the coherent interconnects or HPE's Memory-Driven Computing have been fully built.
When they do, they will require new hardware at all endpoints and new switches. 
Moreover, the interconnects automatically make caches at different endpoints coherent, which could cause performance overhead that is not always necessary (violating \textbf{R3}).

Besides the above \pdm\ works, there are also proposals to include some processing power in between the disaggregated memory layer and the computation layer.
soNUMA~\cite{soNUMA} is a hardware-based solution that scales out NUMA nodes by extending each NUMA node with a hardware unit that services remote memory accesses.
Unlike \sys\ which physically separates \MN{}s from \CN{}s across generic data-center networks, soNUMA still bundles memory and CPU cores, and it is a single-server solution.
Thus, soNUMA works only on a limited scale (violating \textbf{R2}) and is not flexible (violating \textbf{R7}).
MIND~\cite{mind:sosp21}, a concurrent work with \sys, proposes to use a programmable switch for managing coherence directories and memory address mappings between compute nodes and memory nodes.
Unlike \sys\ which adds processing power to every \MN, MIND's single programmable switch has limited hardware resources and could be the bottleneck for both performance and scalability.

{
\begin{figure}[t]
\begin{center}
\footnotesize
\begin{lstlisting}[
numbers=left,
xleftmargin=6.0ex,
xrightmargin=0.1in,
frame=single,
framexleftmargin=15pt
]
/* Alloc one remote page. Define a remote lock */
#define PAGE_SIZE (1<<22)
void *remote_addr = ralloc(PAGE_SIZE);
ras_lock lock;

/* Acquire lock to enter critical section.
   Do two AYSNC writes then poll completion. */
void thread1(void *) {
 rlock(lock);
 e[0]=rwrite(remote_addr, local_wbuf1,len, ASYNC);
 e[1]=rwrite(remote_addr+len, local_wbuf2,len, ASYNC);
 runlock(lock);
 rpoll(e, 2);
}

/* Synchronously read from remote */
void thread2(void *) {
 rlock(lock); 
 rread(remote_addr, local_rbuf, len, SYNC);
 runlock(lock); 
}
\end{lstlisting}
\mycaption{fig-code-eg}{Example of Using \sys.}
{
}
\end{center}
\end{figure}
}

\section{\sys\ Overview}
\label{sec:hdm}

\sys\ co-designs software with hardware, \CN{}s with \MN{}s, and network stack with virtual memory system, 
so that at the \MN{}, the entire data path is handled in hardware with high throughput, low (tail) latency, and minimal hardware resources. 
This section gives an overview of \sys's interface and architecture (Figure~\ref{fig-arch}).


\subsection{\sys\ Interface}
\label{sec:abstraction}

Similar to recent \md\ proposals~\cite{AIFM,sebastian-hotcloud20}, our current implementation adopts a non-transparent interface where
applications (running at \CN{}s) allocate and access disaggregated memory via explicit API calls. Doing so gives users opportunities to perform application-specific performance optimizations. 
By design, \sys’s APIs can also be called by a runtime like the AIFM runtime~\cite{AIFM} or by the kernel/hardware at \CN\ like LegoOS' pComponent~\cite{Shan18-OSDI} to support a transparent interface and allow the use of unmodified user applications.
We leave such extension to future work.

Apart from the regular (local) virtual memory address space, each process has a separate {\em \textbf{R}emote virtual memory \textbf{A}ddress \textbf{S}pace} ({\em \rspace} for short).
Each application process has a unique global {\em PID} across all \CN{}s which is assigned by \sys\ when the application starts.
Overall, programming in \rspace\ is similar to traditional multi-threaded programming except that memory read and write are explicit and that processes running on different \CN{}s can share memory in the same \rspace.
Figure~\ref{fig-code-eg} illustrates the usage of \sys\ with a simple example.

An application process can perform a set of virtual memory operations in its \rspace,
including \alloc, \sysfree, \sysread, \syswrite, 
and a set of atomic and synchronization primitives (\eg, \syslock, \sysunlock, \fence).
\alloc\ works like \texttt{malloc} and returns a VA in \rspace. \sysread\ and \syswrite\ can then be issued to any allocated VAs.
As with the traditional virtual memory interface, allocation and access in \rspace\ are in byte granularity.
We offer {\em synchronous} and {\em asynchronous} options for \alloc, \sysfree, \sysread, and \syswrite.

\ulinebfpara{Intra-thread request ordering.}
Within a thread, synchronous APIs follow strict ordering.
An application thread that calls a synchronous API blocks until it gets the result.
Asynchronous APIs are non-blocking. A calling thread proceeds after calling an asynchronous API and later calls \poll\ to get the result. 
Asynchronous APIs follow a release order.
Specifically, asynchronous APIs may be executed out of order as long as
1) all asynchronous operations before a \release\ complete before the \release\ returns,
and 2) \release\ operations are strictly ordered.
On top of this release order, 
we guarantee that there is no concurrent asynchronous operations with dependencies (Write-After-Read, Read-After-Write, Write-After-Write) and target the same page.
The resulting memory consistency level is the same as architecture like ARMv8~\cite{ARMv8}.
In addition, we also ensure consistency between metadata and data operations, by ensuring that potentially conflicting operations execute synchronously in the program order. For example, if there is an ongoing \sysfree\ request to a VA, no read or write to it can start until the \sysfree\ finishes.
Finally, failed or unresponsive requests are transparently retried, and they follow the same ordering guarantees.

{
\begin{figure}[t!]
\begin{center}
\centerline{\includegraphics[width=1.0\columnwidth]{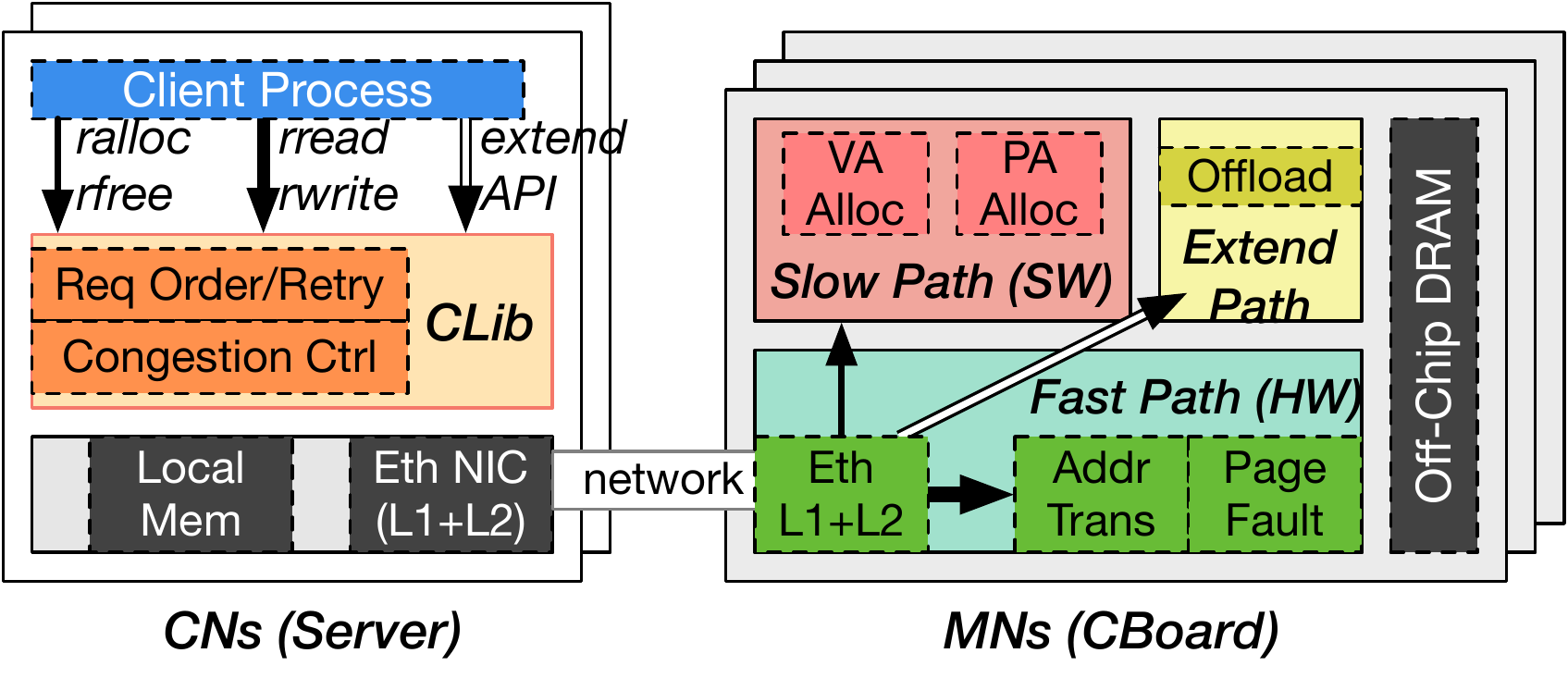}}
\mycaption{fig-arch}{\sys\ Architecture.}
{
}
\end{center}
\end{figure}
}

\ulinebfpara{Thread synchronization and data coherence.}
Threads and processes can share data even when they are not on the same \CN.
Similar to traditional concurrent programming, \sys\ threads can use synchronization primitives to build critical sections (\eg, with \syslock) 
and other semantics (\eg, flushing all requests with \fence).

An application can choose to cache data read from \sysread\ at the \CN\ (\eg, by maintaining \texttt{local\_rbuf} in the code example).
Different processes sharing data in a \rspace\ can have their own cached copies at different \CN{}s.
Similar to ~\cite{Shan18-OSDI}, \sys\ does not make these cached copies coherent automatically and lets applications choose their own coherence
protocols.
We made this deliberate decision because automatic cache coherence on every read/write would incur  high performance overhead with commodity Ethernet infrastructure
and application semantics could reduce this overhead.

\subsection{\sys\ Architecture}

In \sys\ (Figure~\ref{fig-arch}), \CN{}s are regular servers each equipped with a regular Ethernet NIC and connected to a top-of-rack (ToR) switch.
\MN{}s are our customized devices directly connected to a ToR switch.
Applications run at \CN{}s on top of our user-space library called {\em \syslib}.
It is in charge of request ordering, request retry, congestion, and incast control. 

By design, an \MN\ in \sys\ is a \sysboard\ consisting of an ASIC which runs the hardware logic for all data accesses (we call it the {\em fast path} and prototyped it with FPGA),
an ARM processor which runs software for handling metadata and control operations (\ie, the {\em slow path}),
and an FPGA which hosts application computation offloading (\ie, the {\em extend path}).
An incoming request arrives at the ASIC and travels through standard Ethernet physical and MAC layers 
and a Match-and-Action-Table (MAT) that decides which of the three paths the request should go to based on the request type.
If the request is a data access (fast path), it stays in the ASIC and goes through a hardware-based virtual memory system
that performs three tasks in the same pipeline: address translation, permission checking, and page fault handling (if any).
Afterward, the actual memory access is performed through the memory controller, and the response is formed and sent out through the network stack.
Metadata operations such as memory allocation are sent to the slow path. 
Finally, customized requests with 
offloaded computation are handled in the extend path.

\section{\sys\ Design}
\label{sec:design}

This section presents the design challenges of building a hardware-based \md\ system and our solutions.

\subsection{Design Challenges and Principles}
Building a hardware-based \md\ platform is a previously unexplored area and introduces new challenges mainly because of restrictions of hardware and the unique requirements of \md.

\boldpara{Challenge 1: The hardware should avoid maintaining or processing complex data structures}, because unlike software, hardware has limited resources such as on-chip memory and logic cells.
For example, Linux and many other software systems use trees (\eg, the vma tree) for allocation.
Maintaining and searching a big tree data structure in hardware, however, would require huge on-chip memory and many logic cells to perform the look up operation (or alternatively use fewer resources but suffer from performance loss).

\boldpara{Challenge 2: Data buffers and metadata that the hardware uses should be minimal and have bounded sizes}, so that they can be statically planned and fit into the on-chip memory.
Unfortunately, traditional software approaches 
involve various data buffers and metadata that are large and grow with increasing scale.
For example, today's reliable network transports maintain per-connection sequence numbers and buffer unacknowledged packets for packet ordering and retransmission, and they grow with the number of connections.
Although swapping between on-chip and off-chip memory is possible, doing so would increase both tail latency and hardware logic complexity, especially under large scale.

\boldpara{Challenge 3: The hardware pipeline should be deterministic and smooth},
\ie, it uses a bounded, known number of cycles to process a data unit, and for each cycle, the pipeline can take in one new data unit (from the network).
The former would ensure low tail latency, while the latter would guarantee a throughput that could match network line rate.
Another subtle benefit of a deterministic pipeline is that we can know the maximum time a data unit stays at \MN,
which could help bound the size of certain buffers (\eg, \S\ref{sec:ordering}).
However, many traditional hardware solutions are not designed to be deterministic or smooth, and we cannot directly adapt their approaches.
For example, traditional CPU pipelines could have stalls because of data hazards and have non-deterministic latency to handle memory instructions.

To confront these challenges, we took a clean-slate approach by designing \sys's virtual memory system and network system with the following principles that all aim to eliminate state in hardware or bound their performance and space overhead.

\boldpara{Principle 1: Avoid state whenever possible.}
Not all state in server-based solutions is necessary if we could redesign the hardware.
For example, we get rid of RDMA's MR indirection and its metadata altogether
by directly mapping application process' \rspace\ VAs to PAs (instead of to MRs then to PAs).

\boldpara{Principle 2: Moving non-critical operations and state to software and making the hardware fast path deterministic.}
If an operation is non-critical and it involves complex processing logic and/or metadata, our idea is to move it to the software slow path running in an ARM processor.
For example, VA allocation (\alloc) is expected to be a rare operation
because applications know the disaggregated nature and would typically have only a few large allocations during the execution.
Handling \alloc, however, would involve dealing with complex allocation trees.
We thus handle \alloc\ and \sysfree\ in the software slow path.
Furthermore, in order to make the fast path performance deterministic, we {\em decouple} all slow-path tasks from the performance-critical path by {\em asynchronously} performing them in the background.

\boldpara{Principle 3: Shifting functionalities and state to \CN{}s.}
While hardware resources are scarce at \MN{}s, \CN{}s have sufficient memory and processing power, and it is faster to develop functionalities in \CN\ software.
A viable solution is to shift state and functionalities from \MN{}s to \CN{}s.
The key question here is how much and what to shift.
Our strategy is to shift functionalities to \CN{}s only if doing so 1) could largely reduce hardware resource consumption at \MN{}s, 2) does not slow down common-case foreground data operations, 3) does not sacrifice security guarantees, and 4) adds bounded memory space and CPU cycle overheads to \CN{}s.
As a tradeoff, the shift may result in certain uncommon operations (\eg, handling a failed request) being slower.

\boldpara{Principle 4: Making off-chip data structures efficient and scalable.}
Principles 1 to 3 allow us to reduce \MN\ hardware to only the most essential functionalities and state. 
We store the remaining state in off-chip memory and cache a fixed amount of them in on-chip memory.
Different from most caching solutions, our focus is to make the access to off-chip data structure fast and scalable,
\ie, all cache misses have bounded latency regardless of the number of client processes accessing an \MN\ or the amount of physical memory the \MN\ hosts.

\boldpara{Principle 5: Making the hardware fast path smooth by treating each data unit independently at \MN.}
If data units have dependencies (\eg, must be executed in a certain order), then the fast path cannot always execute a data unit when receiving it.
To handle one data unit per cycle and reach network line rate, we make each data unit independent by including all the information needed to process a unit in it and by allowing \MN{}s to execute data units in any order that they arrive.
To deliver our consistency guarantees, we opt for enforcing request ordering at \CN{}s before sending them out.

The rest of this section presents how we follow these principles to design \sys's three main functionalities: memory address translation and protection, page fault handling, and networking. We also briefly discuss our offloading support.

\subsection{Scalable, Fast Address Translation}
\label{sec:addr-trans}
Similar to traditional virtual memory systems, we use fix-size pages as address allocation and translation unit, while data accesses are in the granularity of byte.
Despite the similarity in the goal of address translation,
the radix-tree-style, per-address space page table design used by all current architectures~\cite{ecuckoo-asplos20} does not fit \md\ for two reasons.
First, each request from the network could be from a different client process. If each process has its own page table, \MN\ would need to cache and look up many page table roots, causing additional overhead.
Second, a multi-level page table design requires multiple DRAM accesses when there is a translation lookaside buffer (TLB) miss~\cite{hashpgtable-sigmetrics16}.
TLB misses will be much more common in a \md\ environment, since with more applications sharing an \MN, the total working set size is much bigger than that in a single-server setting, while the TLB size in an \MN\ will be similar or even smaller than a single server's TLB (for cost concerns). To make matters worse, each DRAM access is more costly for systems like RDMA NIC which has to cross the PCIe bus to access the page table in main memory~\cite{Pythia,pcie-sigcomm}.

\ulinebfpara{Flat, single page table design (Principle 4).}~~
We propose a new {\em overflow-free} hash-based page table design that sets the total page table size according to the physical memory size 
and bounds \textit{address translation to at most one DRAM access}.
Specifically, we store {\em all} page table entries (PTEs) from {\em all} processes in a single hash table 
whose size is proportional to the physical memory size of an \MN. 
The location of this page table is fixed in the off-chip DRAM and is known by the fast path address translation unit, thus avoiding any lookups.
As we anticipate applications to allocate big chunks of VAs in their \rspace, we use huge pages and support a configurable set of page sizes.
With the default 4\MB\ page size, the hash table consumes only 0.4\%\ of the physical memory.

{
\begin{figure}[t]
\begin{center}
\centerline{\includegraphics[width=\columnwidth]{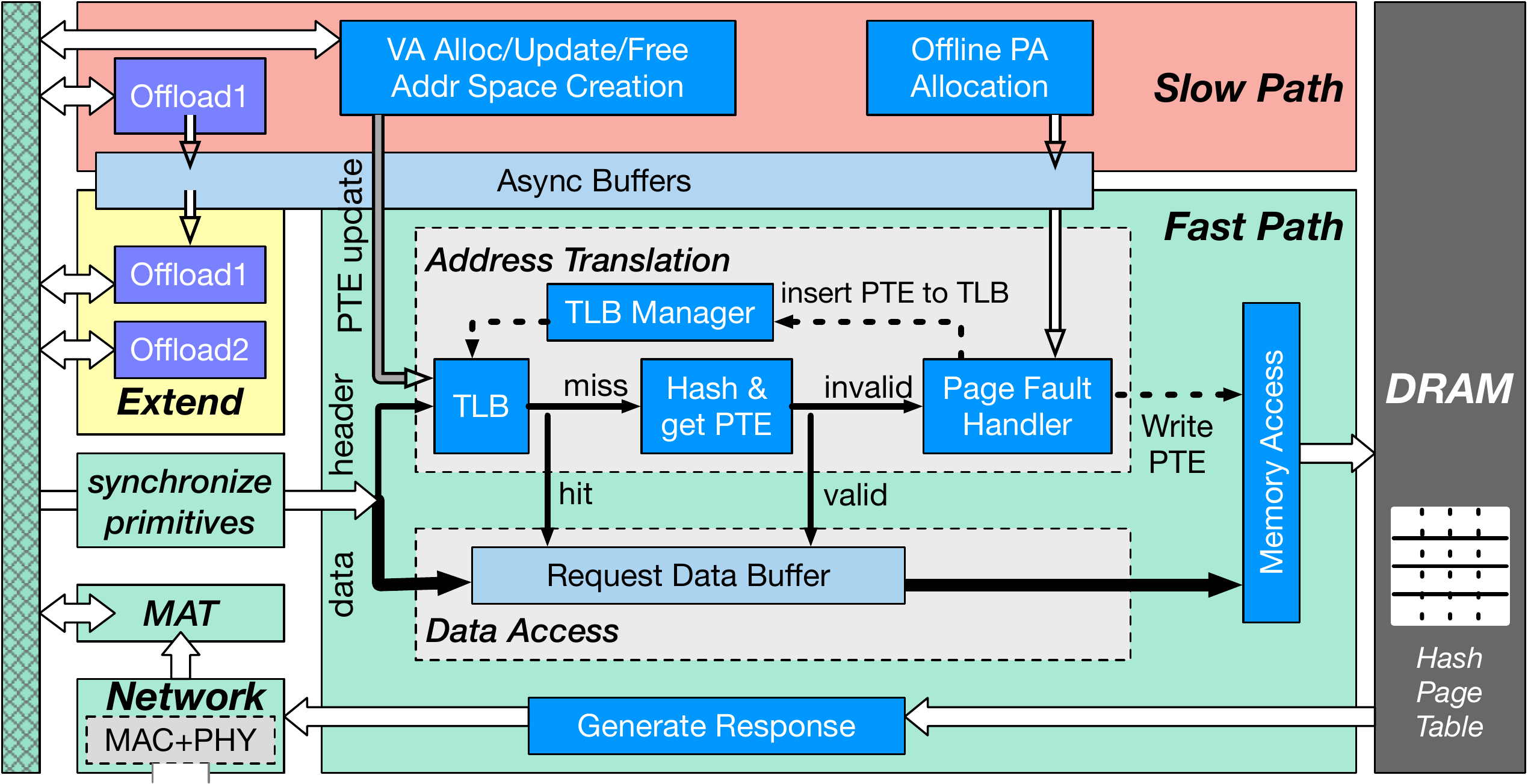}}
\mycaption{fig-coremem}{\sysboard\ Design.}
{
Green, yellow, and red areas are anticipated to be built with 
ASIC, FPGA, and low-power cores.
}
\end{center}
\end{figure}
}

The hash value of a VA and its PID is used as the index to determine which hash bucket the corresponding PTE goes to.
Each hash bucket has a fixed number of ($K$) slots.
To access the page table, we always fetch
the entire bucket including all $K$ slots in a single DRAM access.

A well-known problem with hash-based page table design is hash collisions that could overflow a bucket.
Existing hash-based page table designs rely on collision chaining~\cite{TransCache-isca10} or open addressing~\cite{hashpgtable-sigmetrics16} to handle overflows, both require multiple DRAM accesses or even costly software intervention.
In order to bound address translation to at most one DRAM access, we use a novel technique to avoid hash overflows at \textit{VA allocation time}.

\ulinebfpara{VA allocation (Principle 2).}~~
The slow path software handles \alloc\ requests and allocates VA.
The software allocator maintains a per-process VA allocation tree that records allocated VA ranges and permissions, similar to the Linux vma tree~\cite{linux-rb-vma}.
To allocate size $k$ of VAs, it first finds an available address range of size $k$ in the tree.
It then calculates the hash values of the virtual pages in this address range
and checks if inserting them to the page table would cause any hash overflow. 
If so, it 
does another search for available VAs.
These steps repeat until it finds a valid VA range that does not cause hash overflow.

Our design trades potential retry overhead at allocation time (at the slow path) for better run-time performance and simpler hardware design (at the fast path).
This overhead is manageable because
1) each retry takes only a few microseconds with our implementation (\S\ref{sec:impl}),
2) we employ huge pages, which means fewer pages need to be allocated, 
3) we choose a hash function that has very low collision rate~\cite{lookup3-wiki},
and 4) we set the page table to have extra slots (2\x\ by default) which absorbs most overflows.
We find no conflicts when memory is below half utilized and has only up to 60 retries when memory is close to full (Figure~\ref{fig-alloc-conflict}).


\ulinebfpara{TLB.}~~
\sys\ implements a TLB in a fix-sized on-chip memory area and looks it up using content-addressable-memory in the fast path.
On a TLB miss, the fast path fetches the PTE from off-chip memory and inserts it to the TLB by replacing an existing TLB entry with the LRU policy.
When updating a PTE, the fast path also updates the TLB, in a way that ensures the consistency of inflight operations.

\ulinebfpara{Limitation.}~~
A downside of our overflow-free VA allocation design is that it cannot guarantee that a specific VA can be inserted into the page table. This is not a problem for regular VA allocation but could be problematic for allocations that require a fixed VA (\eg, \texttt{mmap(MAP\_FIXED})). 
Currently, \sys\ finds a new VA range if the user-specified range cannot be inserted into the page table. Applications that must map at fixed VAs (\eg, libraries) will need to use \CN-local memory.

\subsection{Low-Tail-Latency Page Fault Handling}

A key reason to disaggregate memory is to consolidate memory usages on less DRAM so that memory utilization is higher and the total monetary cost is lower (\textbf{R1}). Thus, remote memory space is desired to run close to full capacity, and we allow memory over-commitment at an \MN, necessitating page fault handling. Meanwhile, applications like JVM-based ones allocate a large heap memory space at the startup time and then slowly use it to allocate smaller objects~\cite{ali-trace}. Similarly, many existing far-memory systems~\cite{Tsai20-ATC,AIFM,FaRM} allocate a big chunk of remote memory and then use different parts of it for smaller objects to avoid frequently triggering the slow remote allocation operation.
In these cases, it is desirable for a \md\ system to delay the allocation of physical memory to when the memory is actually used (\ie, {\em on-demand} allocation) or to ``reshape'' memory~\cite{cliquemap-sigcomm21} during runtime, necessitating page fault handling.

Page faults are traditionally signaled by the hardware and handled by the OS. 
This is a slow process because of the costly interrupt and kernel-trapping flow.
For example, a remote page fault via RDMA costs 16.8\ms\ from our experiments using Mellanox ConnectX-4.
To avoid page faults, most RDMA-based systems pre-allocate big chunks of physical memory and pin them physically.
However, doing so results in memory wastes and makes it hard for an \MN\ to pack more applications, violating \textbf{R1} and \textbf{R2}.

We propose to {\em handle page faults in hardware and with bounded latency}\textemdash a {\em constant three cycles} to be more specific with our implementation of \sysboard.
Handling initial-access faults in hardware is challenging, as initial accesses require PA allocation, which is a slow operation that involves manipulating complex data structures.
Thus, we handle PA allocation in the slow path (\textbf{Challenge 1}).
However, if the fast-path page fault handler has to wait for the slow path to generate a PA for each page fault,
it will slow down the data plane.

To solve this problem, we propose an asynchronous design to shift PA allocation off the performance-critical path (\textbf{Principle 2}).
Specifically, we maintain a set of {\em free physical page numbers} in an {\em async buffer},
which the ARM continuously fulfills by finding free physical page addresses and reserving them without actually using the pages. 
During a page fault, the page fault handler simply fetches a pre-allocated physical page address. 
Note that even though a single PA allocation operation has a non-trivial delay, 
the throughput of generating PAs and filling the async buffer is higher than network line rate.
Thus, the fast path can always find free PAs in the async buffer in time.
After getting a PA from the async buffer and establishing a valid PTE, 
the page fault handler performs three tasks in parallel: 
writing the PTE to the off-chip page table, inserting the PTE to the TLB,
and continuing the original faulting request.
This parallel design hides the performance overhead of the first two tasks, allowing foreground requests to proceed immediately.

A recent work~\cite{lee-isca20} also handles page faults in hardware. 
Its focus is on the complex interaction with kernel and storage devices, and it is a simulation-only work. \sys\ uses a different design for handling page faults in hardware with the goal of low tail latency, and we built it in FPGA.

\ulinebfpara{Putting the virtual memory system together.}~~
We illustrate how \sysboard{}'s virtual memory system works using a simple example of allocating some memory and writing to it.
The first step (\alloc) is handled by the slow path, which allocates a VA range by finding an available set of slots in the hash page table.
The slow path forwards the new PTEs to the fast path, which inserts them to the page table.
At this point, the PTEs are invalid.
This VA range is returned to the client.
When the client performs the first write, the request goes to the fast path.
There will be a TLB miss, followed by a fetch of the PTE.
Since the PTE is invalid, the page fault handler will be triggered,
which fetches a free PA from the async buffer and establishes the valid PTE.
It will then execute the write, update the page table, and insert the PTE to TLB.

\subsection{Asymmetric Network Tailored for \md}
\label{sec:network}
With large amounts of research and development efforts, today's data-center network systems are highly optimized in their performance.
Our goal of \sys's network system is unique and fits \md's requirements\textemdash minimizing the network stack's hardware resource consumption at \MN{}s and achieving great scalability while maintaining similar performance as today's fast network.
Traditional software-based reliable transports like Linux TCP incurs high performance overhead.
Today's hardware-based reliable transports like RDMA are fast, but they require a fair amount of on-chip memory to maintain state, \eg, per-connection sequence numbers, congestion state~\cite{TONIC}, and bitmaps~\cite{IRN,MELO-APNet}, not meeting our low-cost goal.

Our insight is that different from general-purpose network communication where each endpoint can be both the sender (requester) and the receiver (responder) that exchange general-purpose messages,
\MN{}s only respond to requests sent by \CN{}s (except for memory migration from one \MN\ to another \MN\ (\S\ref{sec:dist}), in which case we use another simple protocol to achieve the similar goal).
Moreover, these requests are all memory-related operations that have their specific properties.
With these insights, we design a new network system with two main ideas.
Our first idea is to maintain transport logic, state, and data buffers only at \CN{}s,
essentially making \MN{}s ``transportless'' (\textbf{Principle 3}). 
Our second idea is to relax the reliability of the transport and instead enforce ordering and loss recovery at the memory request level, so that \MN{}s' hardware pipeline can process data units as soon as they arrive (\textbf{Principle 5}).

With these ideas, we implemented a transport in \syslib\ at \CN{}s. \syslib\ bypasses the kernel to directly issue raw Ethernet requests to an Ethernet NIC.
\CN{}s use regular, commodity Ethernet NICs and regular Ethernet switches to connect to \MN{}s.
\MN{}s include only standard Ethernet physical, link, and network layers and a slim layer for handling corner-case requests (\S\ref{sec:ordering}).
We now describe our detailed design.


\ulinebfpara{Removing connections with request-response semantics.}
Connections (\ie, QPs) are a major scalability issue with RDMA.
Similar to recent works~\cite{Homa,1RMA}, we make our network system connection-less using request-response pairs.
Applications running at \CN{}s directly initiate \sys\ APIs to an \MN\ without any connections.
\syslib\ assigns a unique request ID to each request. The \MN\ attaches the same request ID when sending the response back. \syslib\ uses responses as ACKs and matches a response with an outstanding request using the request ID.
Neither \CN{}s nor \MN{}s send ACKs.

\ulinebfpara{Lifting reliability to the memory request level.}
Instead of triggering a retransmission protocol for every lost/corrupted packet at the transport layer, 
\syslib\ retries the entire memory request if any packet is lost or corrupted in the sending or the receiving direction.
On the receiving path, \MN{}'s network stack only checks a packet's integrity at the link layer. If a packet is corrupted, the \MN\ immediately sends a NACK to the sender \CN.
\syslib\ retries a memory request if one of three situations happens: a NACK is received, the response from \MN\ is corrupted, or no response is received within a \texttt{TIMEOUT} period.
In addition to lifting retransmission from transport to the request level, we also lift ordering to the memory request level
and allow out-of-order packet delivery (see details in \S\ref{sec:ordering}). 

\ulinebfpara{\CN-managed congestion and incast control.}
Our goal of controlling congestion in the network and handling incast that can happen both at a \CN\ and an \MN\ is to minimize state at \MN.
To this end, we build the entire congestion and incast control at the \CN\ in the \syslib.
To control congestion, \syslib\ adopts a simple delay-based, reactive policy that uses end-to-end RTT delay as the congestion signal, similar to recent sender-managed, delay-based mechanisms~\cite{mittal2015timely,swift-sigcomm,1RMA}.
Each \CN\ maintains one congestion window, \textit{cwnd}, per \MN\ 
that controls the maximum number of outstanding requests that can be made to the \MN\ from this \CN.
We adjust \textit{cwnd} based on measured delay using a standard Additive Increase Multiplicative Decrease (AIMD) algorithm.

To handle incast to a \CN, we exploit the fact that the \CN{} knows the sizes of expected responses for the requests that it sends out and that responses are the major incoming traffic to it.
Each \syslib\ maintains one incast window, \textit{iwnd}, which controls the maximum bytes of expected responses. \syslib\ sends a request only when both \textit{cwnd} and \textit{iwnd} have room.

Handling incast to an \MN\ is more challenging, as we cannot throttle incoming traffic at the \MN\ side or would otherwise maintain state at \MN{}s.
To have \CN{}s handle incast to \MN{}s, we draw inspiration from Swift~\cite{swift-sigcomm} by allowing \textit{cwnd} to fall below one packet when long delay is observed at a \CN. For example, a \textit{cwnd} of 0.1 means that the \CN\ can only send a packet within 10 RTTs.
Essentially, this situation happens when the network between a \CN\ and an \MN\ is really congested, and the only way is to slow the sending speed.

\subsection{Request Ordering and Data Consistency}
\label{sec:ordering}

As explained in \S\ref{sec:abstraction}, \sys\ supports both synchronous and asynchronous remote memory APIs, with the former following a sequential, one-at-a-time order in a thread and the latter following a release order in a thread.
Furthermore, \sys\ provides synchronization primitives for inter-thread consistency.
We now discuss how \sys\ achieves these correctness guarantees by presenting our mechanisms for handling intra-request intra-thread ordering, inter-request intra-thread ordering, inter-thread consistency, and retries.
At the end, we will provide the rationales behind our design.

One difficulty in designing the request ordering and consistency mechanisms is our relaxed network ordering guarantees, 
which we adopt to minimize the hardware resource consumption for the network layer at \MN{}s (\S\ref{sec:network}).
On an asynchronous network, it is generally hard to guarantee any type of request ordering when there can be multiple outstanding requests (either multiple threads accessing shared memory or a single thread issuing multiple asynchronous APIs). It is even harder for \sys\ because we aim to make \MN\ stateless as much as possible.
Our general approaches are 1) using \CN{}s to ensure that no two concurrently outstanding requests are dependent on each other, and 2) using \MN{}s to ensure that every user request is only executed once even in the event of retries.




\ulinebfpara{Allowing intra-request packet re-ordering (T1).}
A request or a response in \sys\ can contain multiple link-layer packets. 
Enforcing packet ordering above the link layer normally requires maintaining state (\eg, packet sequence ID) at both the sender and the receiver.
To avoid maintaining such state at \MN{}s,
our approach is to deal with packet reordering only at \CN{}s in \syslib\ (\textbf{Principle 3}).
Specifically, \syslib\ splits a request that is bigger than link-layer maximum transmission unit (MTU) into several link-layer packets
and attaches a \sys\ header to each packet, which includes sender-receiver addresses, a request ID, and request type.
This enables the \MN{} to treat each packet independently (\textbf{Principle 5}).
It executes packets as soon as they arrive, even if they are not in the sending order.
This out-of-order data placement semantic is in line with RDMA specification~\cite{IRN}. 
Note that only write requests will be bigger than MTU, and the order of data writing within a write request does not affect correctness as long as proper {\em inter-request} ordering is followed.
When a \CN\ receives multiple link-layer packets belonging to the same request response, 
\syslib\ reassembles them before delivering them to the application.

\ulinebfpara{Enforcing intra-thread inter-request ordering at \CN\ (T2).}
Since only one synchronous request can be outstanding in a thread, there cannot be any inter-request reordering problem.
On the other hand, there can be multiple outstanding asynchronous requests.
Our provided consistency level disallows concurrent asynchronous requests that are dependent on each other (WAW, RAW, or WAR).
In addition, all requests must complete before \release.


We enforce these ordering requirements at \CN{}s in \syslib\ instead of at \MN{}s (\textbf{Principle 3}) for two reasons.
First, enforcing ordering at \MN{}s requires more on-chip memory and complex logic in hardware.
Second, even if we enforce ordering at \MN{}s, network reordering would still break end-to-end ordering guarantees.

Specifically, \syslib\ keeps track of all inflight requests and matches every new request's virtual page number (VPN) to the inflight ones'. 
If a WAR, RAW, or WAW dependency is detected, \syslib\ blocks the new request until the conflicting request finishes.
When \syslib\ sees a \release\ operation, it waits until all inflight requests return or time out.
We currently track dependencies at the page granularity mainly to reduce tracking complexity and metadata overhead. The downside is that false dependencies could happen (\eg, two accesses to the same page but different addresses).
False dependencies could be reduced by dynamically adapting the tracking granularity if application access patterns are tracked\textemdash we leave this improvement for future work.

\ulinebfpara{Inter-thread/process consistency (T3).}
Multi-threaded or multi-process concurrent programming on \sys\ could use the synchronization primitives \sys\ provides to ensure data consistency (\S\ref{sec:abstraction}).
We implemented all synchronization primitives like \syslock\ and \fence\ at \MN,
because they need to work across threads and processes that possibly reside on different \CN{}s.
Before a request enters either the fast or the slow paths, 
\MN\ checks if it is a synchronization primitive.
For primitives like \syslock\ that internally is implemented using atomic operations like \texttt{TAS}, \MN\ blocks future atomic operations until the current one completes.
For \fence, \MN\ blocks all future requests until all inflight ones complete.
Synchronization primitives are one of the only two cases where \MN\ needs to maintain state.
As these operations are infrequent and each of these operations executes in bounded time, the hardware resources for maintaining their state are minimal and bounded.

\ulinebfpara{Handling retries (T4).}
\syslib\ retries a request after a \texttt{TIMEOUT} period without receiving any response. Potential consistency problems could happen as \sysboard\ could execute a retried write after the data is written by another write request thus undoing this other request's write. Such situations could happen when the original request's response is lost or delayed and/or when the network reorders packets. 
%
We use two techniques to solve this problem.

First, \syslib\ attaches a new request ID to each retry, essentially making it a new request with its own matching response. Together with \syslib's ordering enforcement, it ensures that there is only one outstanding request (or a retry) at any time.
Second, we maintain a small buffer at \MN\ to record the request IDs of recently executed writes and atomic APIs and the results of the atomic APIs. A retry attaches its own request ID and the ID of the failed request. If \MN\ finds a match of the latter in the buffer, it will not execute the request. For atomic APIs, it sends the cached result as the response. We set this buffer's size to be 3$\times$\texttt{TIMEOUT}$\times$\textit{bandwidth}, which is 30\KB\ in our setting. It is one of the only two types of state \MN\ maintains and does not affect the scalability of \MN, since its size is statically associated with the link bandwidth and the \texttt{TIMEOUT} value.
With this size, the \MN\ can ``remember'' an operation long enough for two retries from the \CN. Only when both retries and the original request all fail, the \MN\ will fail to properly handle a future retry. This case is extremely rare~\cite{Homa}, and we report the error to the application, similar to \cite{Kalia14-RDMAKV,1RMA}.

\ulinebfpara{Why T1 to T4?}
We now briefly discuss the rationale behind why we need all T1 to T4 to properly deliver our consistency guarantees. 
First, assume that there is no packet loss or corruption (\ie, no retry) but the network can reorder packets. 
In this case, using T1 and T2 alone is enough to guarantee the proper ordering of \sys\ memory operations, since they guarantee that network reordering will only affect either packets within the same request or requests that are not dependent on each other.
T3 guarantees the correctness of synchronization primitives since the \MN\ is the serialization point and is where these primitives are executed.
Now, consider the case where there are retries.
Because of the asynchronous network, a timed-out request could just be slow and still reach the \MN, either before or after the execution of the retried request. If another request is executed in between the original and the retried requests, inconsistency could happen (\eg, losing the data of this other request if it is a write). The root cause of this problem is that one request can be executed twice when it is retried.
T4 solves this problem by ensuring that the \MN\ only executes a request once even if it is retried.

\subsection{Extension and Offloading Support}
\label{sec:extended}
To avoid network round trips when working with complex data structures and/or performing data-intensive operations,
we extend the core \MN\ to support application computation offloading in the extend path.
Users can write and deploy application offloads both in FPGA and in software (run in the ARM).
To ease the development of offloads, \sys\ offers the same virtual memory interface as the one to applications running at \CN{}s.
Each offload has its own PID and virtual memory address space, and they use the same virtual memory APIs (\S\ref{sec:abstraction}) to access on-board memory. It could also share data with processes running at \CN{}s in the same way that two \CN\ processes share memory.
Finally, an offload’s data and control paths could be split to FPGA and ARM and use the same async-buffer mechanism for communication between them. 
These unique designs made developing computation offloads easier and closer to traditional multi-threaded software programming.

\subsection{Distributed \MN{}s}
\label{sec:dist}
Our discussion so far focused on a single \MN\ (\sysboard).
To more efficiently use remote memory space and to allow one application to use more memory than what one \sysboard\ can offer, we extend the single-\MN\ design to a distributed one with multiple \MN{}s.
Specifically, an application process' \rspace\ can span multiple \MN{}s, and one \MN\ can host multiple \rspace{}s.
We adopt LegoOS' two-level distributed virtual memory management approach to manage distributed \MN{}s in \sys.
A global controller manages \rspace{}s in coarse granularity (assigning 1\GB\ virtual memory regions to different \MN{}s).
Each \MN\ then manages the assigned regions at fine granularity.

The main difference between LegoOS and \sys's distributed memory system is that in \sys, each \MN\ can be over-committed (\ie, allocating more virtual memory than its physical memory size), and when an \MN\ is under memory pressure, it migrates data to another \MN\ that is less pressured (coordinated by the global controller).
The traditional way of providing memory over-commitment is through memory swapping, which could be potentially implemented by swapping memory between \MN{}s. 
However, swapping would cause performance impact on the data path and add complexity to the hardware implementation.
Instead of swapping, we \textit{proactively} migrate a rarely accessed memory region to another \MN\ when an \MN\ is under memory pressure (its free physical memory space is below a threshold).
During migration, we pause all client requests to the region being migrated.
With our 10\Gbps\ experimental board, migrating a 1\GB\ region takes 1.3 second.
Migration happens rarely and, unlike swapping, happens in the background.
Thus, it has little disturbance to foreground application performance.

\section{\sys\ Implementation}
\label{sec:impl}

Apart from challenges discussed in \S\ref{sec:design}, our implementation of \sys\ also needs to overcome several practical challenges, for example, how can different hardware components most efficiently work together in \sysboard, how to minimize software overhead in \syslib. 
This section describes how we implemented \sysboard\ and \syslib, focusing on the new techniques we designed to overcome these challenges.
Currently, \sys\ consists of 24.6K SLOC (excluding computation offloads and third-party IPs).
They include 5.6K SLOC in SpinalHDL~\cite{SpinalHDL} and 2K in C HLS for FPGA hardware, and 17K in C for \syslib\ and ARM software.
We use vendor-supplied interconnect and DDR IPs, and an open-source MAC and PHY network stack~\cite{Corundum-FCCM20}.

\ulinebfpara{\sysboard\ Prototyping.}~~
We prototyped \sysboard\ with a low-cost (\$2495 retail price) Xilinx MPSoC board~\cite{ZCU106} and build the hardware fast path (which is anticipated to be built in ASIC) with FPGA.
All \sys's FPGA modules run at 250\,MHz clock frequency and 512-bit data width.
They all 
achieve an {\em Initiation Interval} ({\em II}) of one
(II is the number of clock cycles between the start time
of consecutive loop iterations, and it decides the maximum
achievable bandwidth). Achieving II of one is not easy and
requires careful pipeline design in all the modules. With II one, our data path can
achieve a maximum of 128\Gbps\ throughput even with just the slower FPGA clock frequency and would be higher with real ASIC implementation.

Our prototyping board consists of a small FPGA with 504K logic cells (LUTs) and 4.75\MB\ FPGA memory (BRAM),
a quad-core ARM Cortex-A53 processor,
two 10\Gbps\ SFP+ ports connected to the FPGA, 
and 2\GB\ of off-chip on-board memory.
This board has several differences from our anticipated real \sysboard:
its network port bandwidth and on-board memory size are both much lower than our target,
and like all FPGA prototypes, its clock frequency is much lower than real ASIC.
Unfortunately, no board on the market offers the combination of small FPGA/ARM (required for low cost) 
and large memory and high-speed network ports. 

Nonetheless, certain features of this board are likely to exist in a real \sysboard,
and these features guide our implementation.
Its ARM processor and the FPGA connect through an interconnect that has high bandwidth (90\GB/s) but high delay (40\mus).
Although better interconnects could be built, crossing ARM and FPGA would inevitably incur non-trivial latency.
With this board, the ARM's access to on-board DRAM is much slower than the FPGA's access because the ARM has to first physically cross the FPGA then to the DRAM.
A better design would connect the ARM directly to the DRAM, 
but it will still be slower for the ARM to access on-board DRAM than its local on-chip memory.


%

To mitigate the problem of slow accesses to on-board DRAM from ARM,
we maintain shadow copies of metadata at ARM's local DRAM.
For example, we store a {\em shadow} version of the page table in ARM's local memory,
so that the control path can read page table content faster.
When the control path needs to perform a virtual memory space allocation, it reads the shadow page table to test if an address would cause an overflow (\S\ref{sec:addr-trans}).
We keep the shadow page table in sync with the real page table by updating both tables when adding, removing, or updating the page table entries.
%

In addition to maintaining shadow metadata, we employ an efficient polling mechanism for ARM/FPGA communication.
We dedicate one ARM core to busy poll an RX ring buffer between ARM and FPGA,
where the FPGA posts tasks for ARM.
This polling thread hands over tasks to other worker threads for task handling 
and post responses to a TX ring buffer.

\sysboard's network stack builds on top of standard, vendor-supplied Ethernet physical and link-layer IPs, with just an additional thin checksum-verify and ack-generation layer on top.
This layer uses much fewer resources compared to a normal RDMA-like stack (\S\ref{sec:results-cost}).
We use lossless Ethernet with Priority Flow Control (PFC) for less packet loss and retransmission. Since PFC has issues like head-of-line blocking~\cite{DCQCN-sigcomm15,hpcc-sigcomm19,alibaba-rdma-nsdi21,IRN}, we rely on our congestion and incast control to avoid triggering PFC as much as possible.

Finally, to assist \sys\ users in building their applications, we implemented a simple software simulator
of \sysboard\ which works with \syslib\ for developers to test their code without the need to run an actual \sysboard.

\ulinebfpara{\syslib\ Implementation.}~~
Even though we optimize the performance of \sysboard, the end-to-end application performance can still be hugely impacted if the host software component (\syslib) is not as fast.
Thus, our \syslib\ implementation aims to provide low-latency performance by adopting several ideas (e.g., data inlining, doorbell batching) from recent low-latency I/O solutions~\cite{ERPC,Kalia14-RDMAKV,Kalia16-ATC,Tsai17-SOSP,Shinjuku,Shenango,demikernel-sosp21}.
We implemented \syslib\ in the user space. 
It has three parts: a user-facing request ordering layer that performs dependency check and ordering of address-conflicting requests,
a transport layer that performs congestion/incast control and request-level retransmission, 
and a low-level device driver layer that interacts with the NIC (similar to DPDK~\cite{DPDK} but simpler).
\syslib\ bypasses kernel and directly issues raw Ethernet requests to the NIC with zero memory copy.
For synchronous APIs, we let the requesting thread poll the NIC for receiving the response right after each request.
For asynchronous APIs, the application thread proceeds with other computations after issuing the request and only busy polls when the program calls \poll.

\section{Building Applications on \sys}
\label{sec:app}

We built five applications on top of \sys, one that uses the basic \sys\ APIs, one that implements and uses a high-level, extended API, and two that offload data processing tasks to \MN{}s, and one that splits computation across \CN{}s and \MN{}s.

\ulinebfpara{Image compression.}
We build a simple image compression/decompression utility that runs purely at \CN.
Each client of the utility (\eg, a Facebook user) has its own collection of photos, 
stored in two arrays at \MN{}s, one for compressed and one for original, both allocated with \alloc.
Because clients' photos need to be protected from each other, we use one process
per client to run the utility.
The utility simply reads a photo from \MN\ using \sysread, compresses/decompresses it,
and writes it back to the other array using \syswrite.
Note that we use compression and decompression as an example of image processing.
These operations could potentially be offloaded to \MN{}s.
However, in reality, there can be many other types of image processing that are more complex and are hard and costly to implement in hardware, necessitating software processing at \CN{}s.
We implemented this utility with 1K C code in 3 developer days.

\ulinebfpara{Radix tree.}
To demonstrate how to build a data structure on \sys\
using \sys's extended API, we built a radix tree with linked lists and pointers.
Data-structure-level systems like AIFM~\cite{AIFM} could follow this example to make simple changes in their libraries to run on \sys.
We first built an extended pointer-chasing functionality in FPGA at the \MN\ which follows pointers in a linked list and performs a value comparison
at each traversed list node. It returns either the node value when there is a match or null when the next pointer becomes null. 
We then expose this functionality to \CN{}s as an extended API.
The software running at \CN\ allocates a big contiguous remote memory space using \alloc\ and uses this space to store radix tree nodes. Nodes in each layer are linked to a list.
To search a radix tree, the \CN\ software goes through each layer of the tree and calls the pointer chasing API until a match is found.
We implemented the radix tree with 300 C code at \CN\ and 150 SpinalHDL code at \sysboard\ in less than one developer day.

\ulinebfpara{Key-value store.}
We built {\em \syskv}, a key-value store that supports concurrent 
create/update/read/delete key-value entries
with atomic write and read committed consistency.
\syskv\ runs at an \MN\ as a computation offloading module.
Users can access it through a key-value interface from multiple \CN{}s.
The \syskv\ module has its own virtual memory address space and uses \sys\ virtual memory APIs to access it.
\syskv\ uses a chained hash table in its virtual memory space for managing the metadata of key-value pairs, and it stores the actual key values at separate locations in the space.
Each hash bucket has a chain of slots. Each slot
contains the virtual addresses of seven key-value pairs.
It also stores a fingerprint for each key-value pair.

To create a new key-value pair, \syskv\ allocates space for
the key-value data with an \alloc\ call and writes the
data with an \syswrite. It then calculates the hash and the fingerprint of the key. 
Afterward, it fetches the last hash slot in the corresponding hash bucket using the hash value. If that
slot is full, \syskv\ allocates another slot using \alloc; otherwise, it just uses the fetched last slot. 
It then inserts the virtual address and fingerprint of the data into the last/new slot.
Finally, it links the current last slot to the new slot if a new one is created.

To perform a read, \syskv\ locates the hash bucket (with the key's hash value) and fetches one slot in the bucket
chain at a time using \sysread. It then compares the fingerprint of the key to the seven entries in the slot. If there is no match, it
fetches the next slot in the bucket. Otherwise, with a matched
entry, it reads the key-value pair using the address stored in that
entry with an \sysread. It then compares the full key and returns
the value if it is a match. Otherwise, it keeps searching the bucket.

The above describes a single-\MN\ \syskv\ system. Another \CN-side load balancer is used to partition key-value pairs into different \MN{}s.
Since all \CN{}s requests of the same partition go to the same \MN\ and \sys\ APIs within an \MN\ are properly ordered, it is fairly easy for \syskv\ to guarantee the atomic-write, read-committed consistency level.

We implemented \syskv\ with 772 SpinalHDL code in 6 developer days.
To evaluate \sys's virtual memory API overhead at \sysboard, we also implemented a key-value store with the same design as \syskv\ but with raw physical memory interface.
This physical-memory-based implementation takes more time to develop and only yields 4\%–12\% latency improvement and 1\%–5\% throughput improvement over \syskv. 

\ulinebfpara{Multi-version object store.}
We built a multi-version object store ({\em \sysmv}) which lets users on \CN{}s create an object, append a new version to an object, 
read a specific version  or the latest version of an object, and delete an object.
Similar to \syskv, \sysmv\ has its own address space.
In the address space, it uses an array to store versions of data for each object, a map to store the mapping from object IDs to the per-object array addresses, and a list to store free object IDs. 
When a new object is created, \sysmv\ allocates a new array (with \alloc) 
and writes the virtual memory address of the array into the object ID map.
Appending a new version to an object simply increases the latest version number
and uses that as an index to the object array for writing the value.
Reading a version simply reads the corresponding element of the array.

\sysmv\ allows concurrent accesses from \CN{}s to an object and guarantees sequential consistency for each object.
Each \sysmv\ user request involves at least two internal
\sys\ operations, some of which include both metadata and
data operations. This compound request pattern makes it tricky
to deal with synchronization problems, as \sysmv\ needs to
ensure that no internal \sys\ operation of a later \sysmv\ request could affect the correctness of an earlier \sysmv\ request. 
Fortunately, both \sys's fast path and slow path guarantee sequential delivery of \sys\ operations. Since \sysmv\ only issues one Clio operation per
clock cycle, the ordering that \sys\ modules guarantee is sufficient to deliver \sysmv’s consistency guarantees. 
We implemented \sysmv\ with 1680 lines of C HLS code in 15 developer days.
 
\ulinebfpara{Simple data analytics.}
Our final example is a simple DataFrame-like data processing application ({\em \sysdf}),
which splits its computation between \CN\ and \MN.
We implement \texttt{select} and \texttt{aggregate} at \MN\ as two offloads,
as offloading them can reduce the amount of data sent over the network.
We keep other operations like \texttt{shuffle} and \texttt{histogram} at \CN.
For the same user, all these modules share the same address space regardless of whether they are at \CN\ or \MN.
Thanks to \sys's support of computation offloading sharing the same address space as computations running at host, \sysdf's implementation is largely simplified and its performance is improved by avoiding data serialization/deserialization.
We implemented \sysdf\ with 202 lines of SpinalHDL code and 170 lines of C interface code in 7 developer days.

{
\begin{figure*}[th]
\begin{minipage}{\figWidthSix}
\begin{center}
\centerline{\includegraphics[width=\columnwidth]{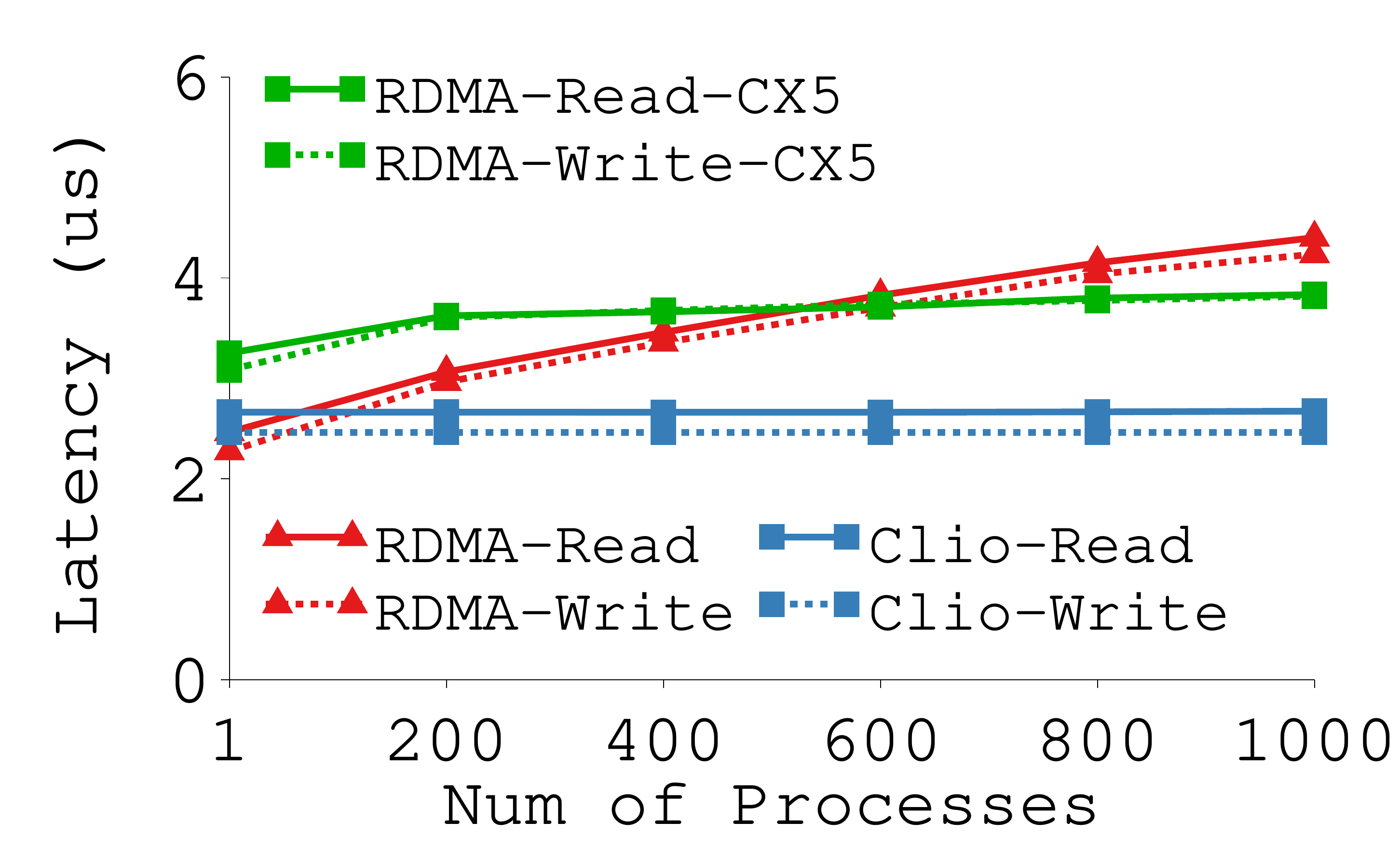}}
\vspace{-0.1in}
\mycaption{fig-conn}{Process (Connection) Scalability.}
{
}
\end{center}
\end{minipage}
\begin{minipage}{\figWidthSix}
\begin{center}
\centerline{\includegraphics[width=\columnwidth]{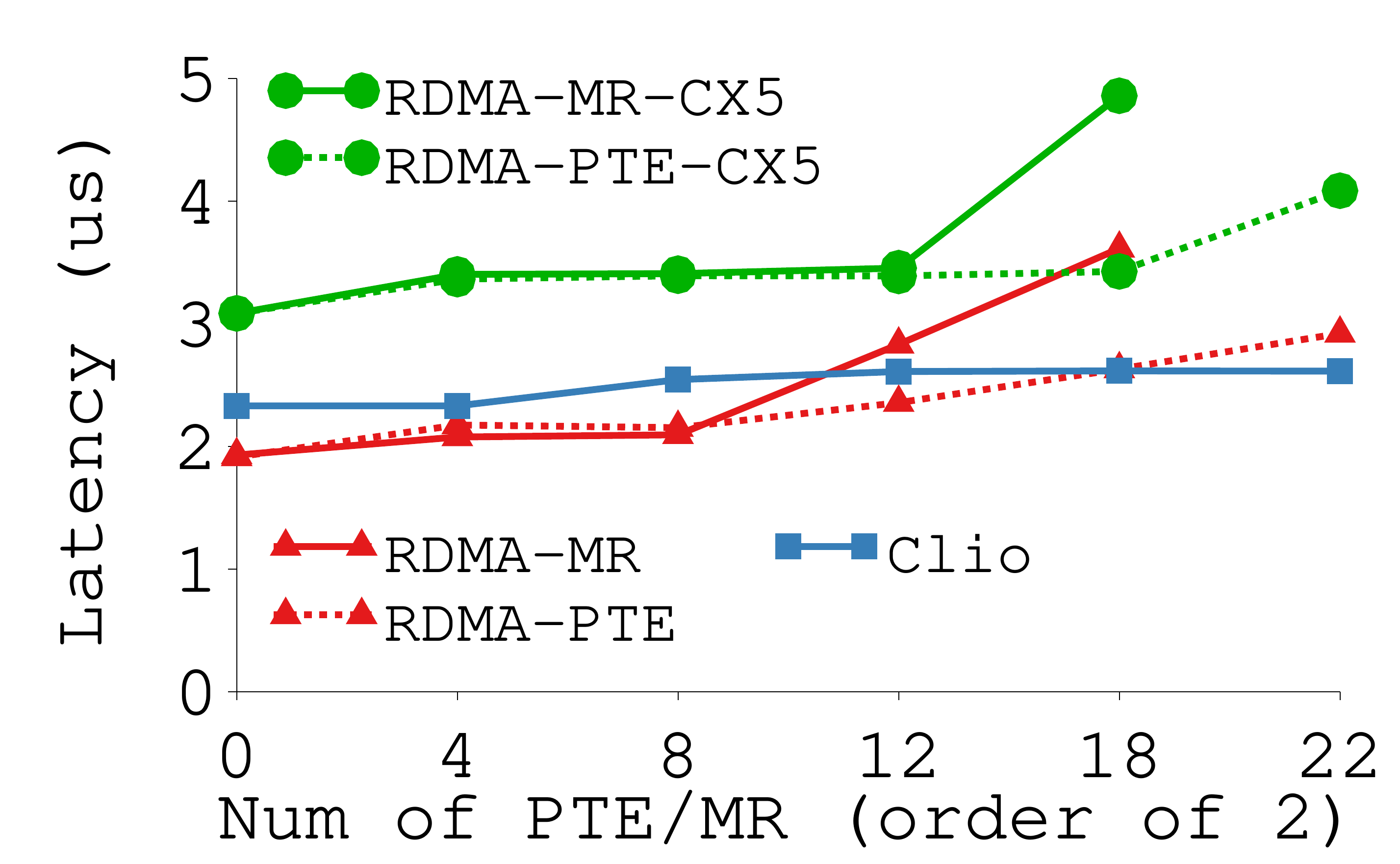}}
\vspace{-0.1in}
\mycaption{fig-pte-mr}{PTE and MR Scalability.}
{
RDMA fails beyond $2^{18}$ MRs. 
}
\end{center}
\end{minipage}
\begin{minipage}{\figWidthSix}
\begin{center}
\centerline{\includegraphics[width=\columnwidth]{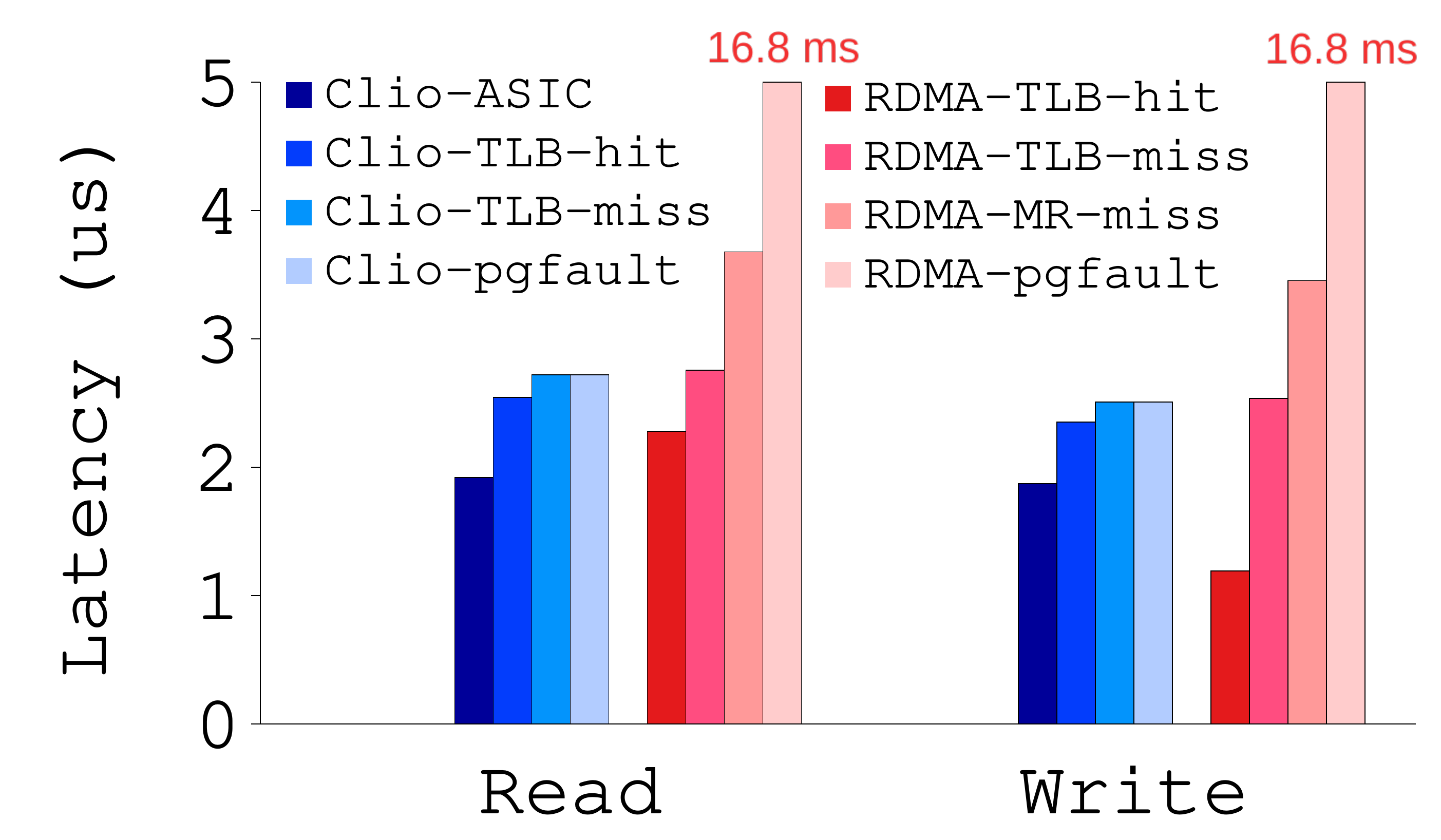}}
\vspace{-0.1in}
\mycaption{fig-miss-hit}{Comparison of TLB Miss and page fault.}
{
\sys-ASIC are projected values of TLB hit.
}
\end{center}
\end{minipage}
\begin{minipage}{\figWidthSix}
\begin{center}
\centerline{\includegraphics[width=\columnwidth]{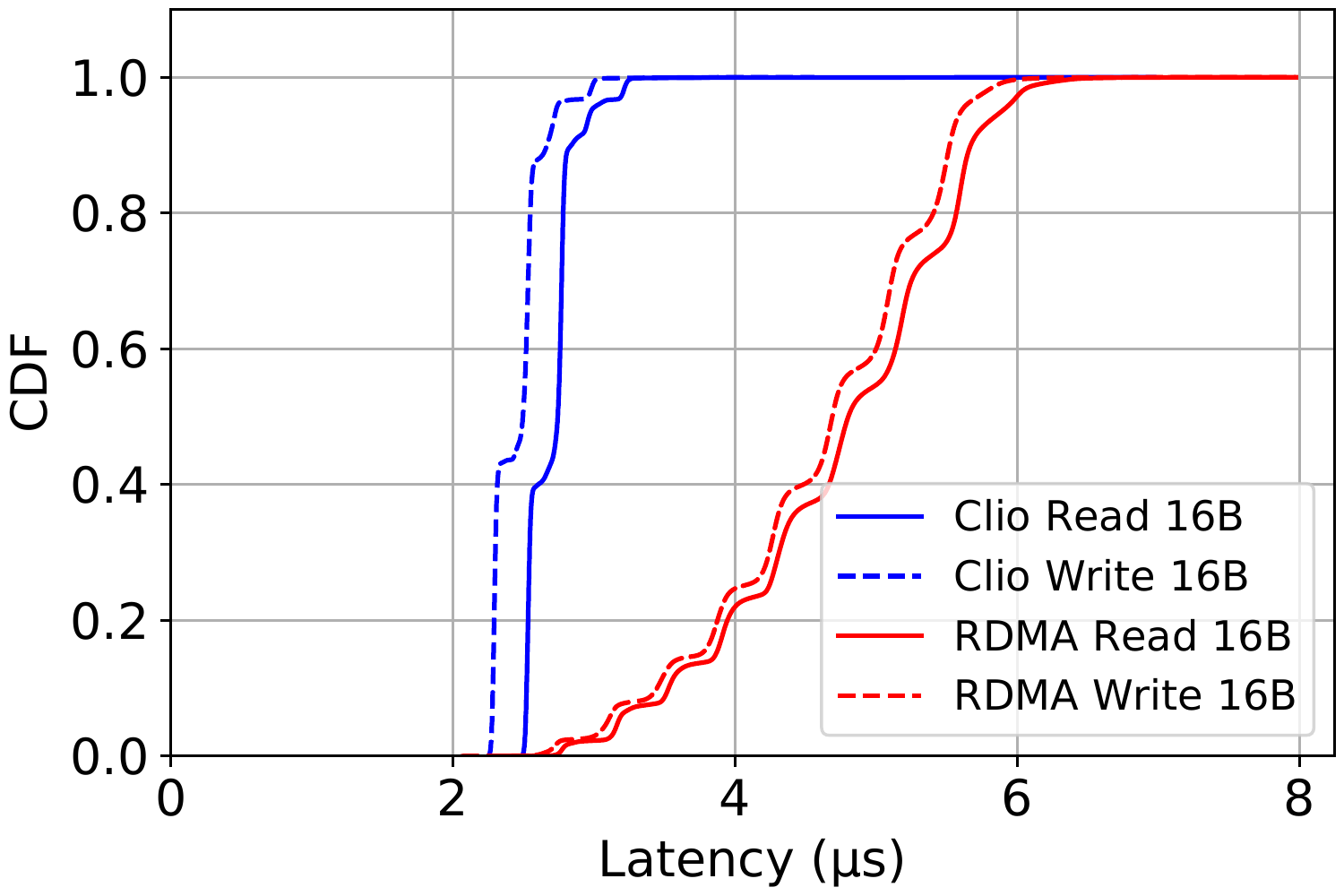}}
\vspace{-0.1in}
\mycaption{fig-tail-latency}{Latency CDF.}
{
}
\end{center}
\end{minipage}
\vspace{-0.1in}
\end{figure*}
}

\section{Evaluation}
\label{sec:results}

{
\begin{figure*}[th]
\begin{minipage}{\figWidthSix}
\begin{center}
\centerline{\includegraphics[width=\columnwidth]{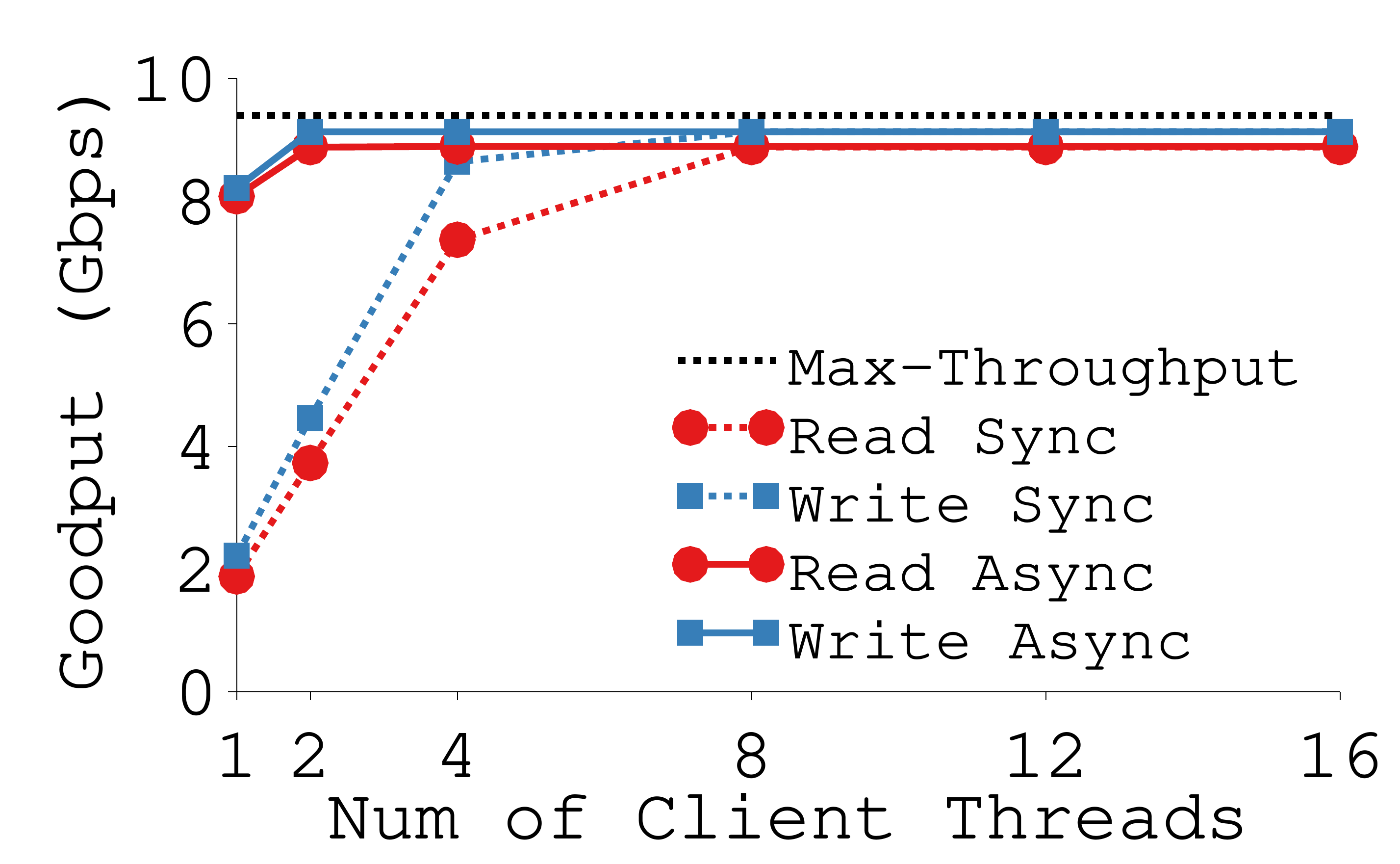}}
\vspace{-0.1in}
\mycaption{fig-read-write-throughput}{End-to-End Goodput.}
{
1\KB\ requests. 
}
\end{center}
\end{minipage}
\begin{minipage}{\figWidthSix}
\begin{center}
\centerline{\includegraphics[width=\columnwidth]{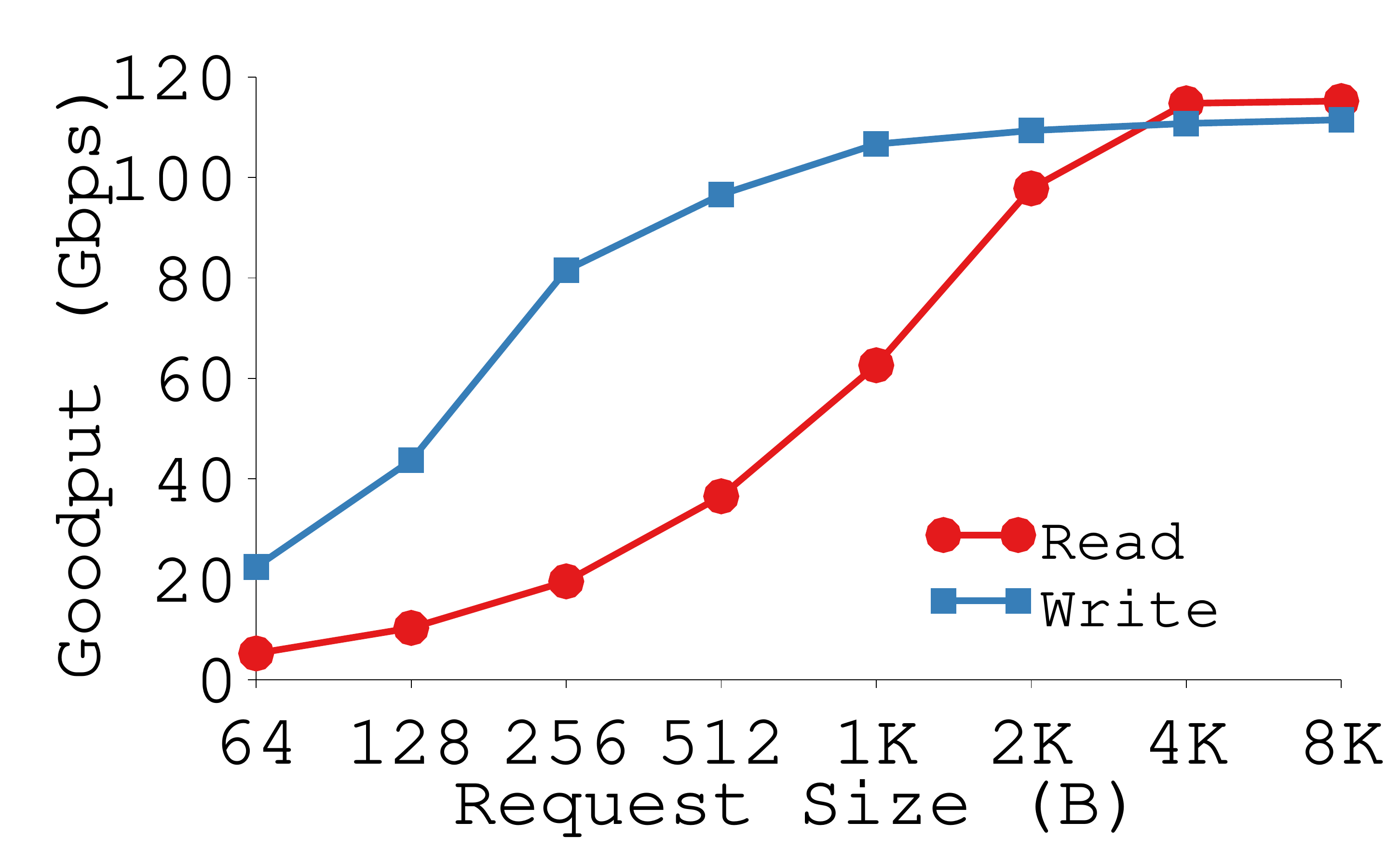}}
\vspace{-0.1in}
\mycaption{fig-onboard-throughput}{On-board Goodput.}
{
FPGA test module generates requests at maximum speed.
}
\end{center}
\end{minipage}
\begin{minipage}{\figWidthSix}
\begin{center}
\centerline{\includegraphics[width=\columnwidth]{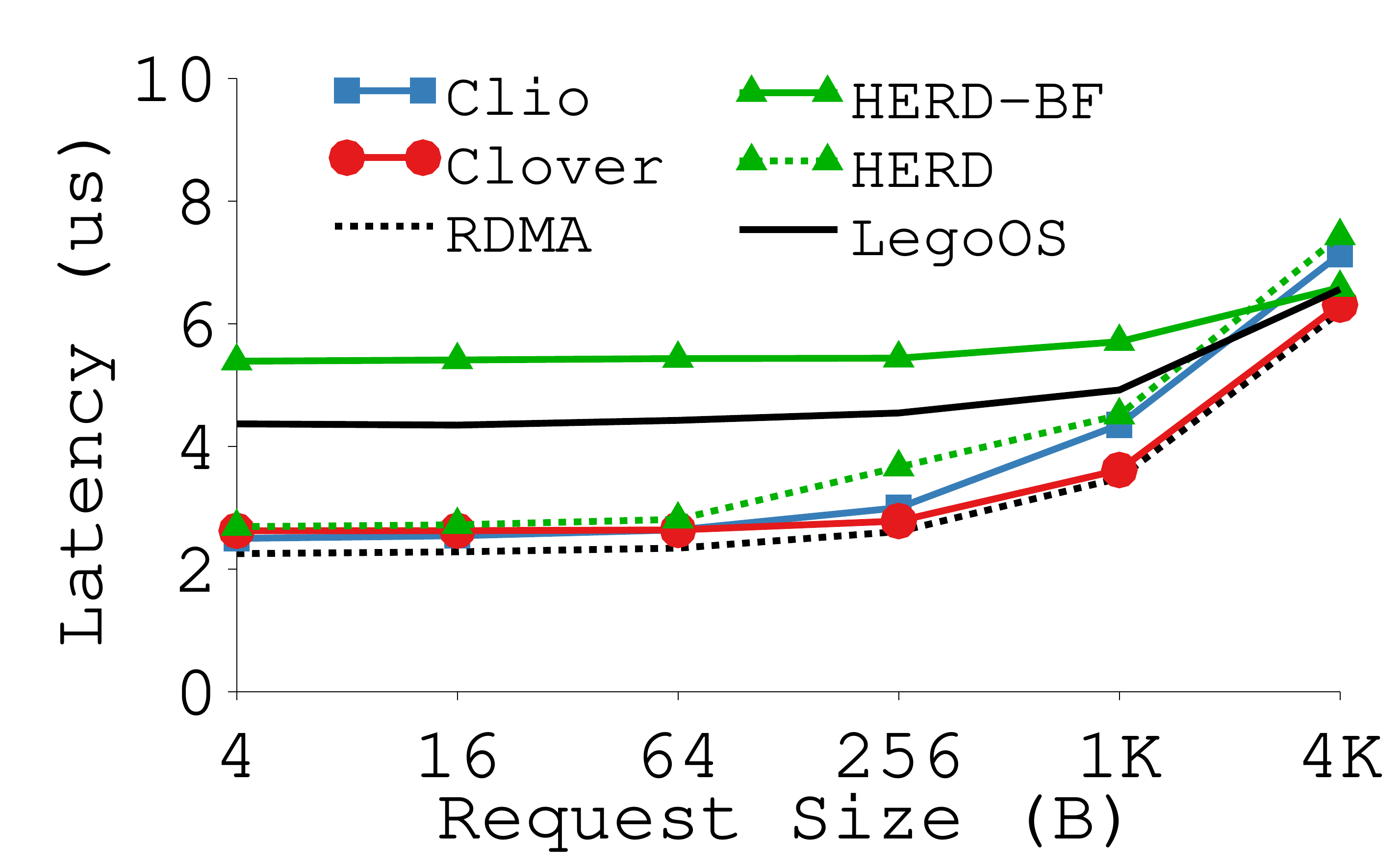}}
\vspace{-0.1in}
\mycaption{fig-read-lat}{Read Latency.}
{
HERD-BF: HERD running on BlueField. 
}
\end{center}
\end{minipage}
\begin{minipage}{\figWidthSix}
\begin{center}
\centerline{\includegraphics[width=\columnwidth]{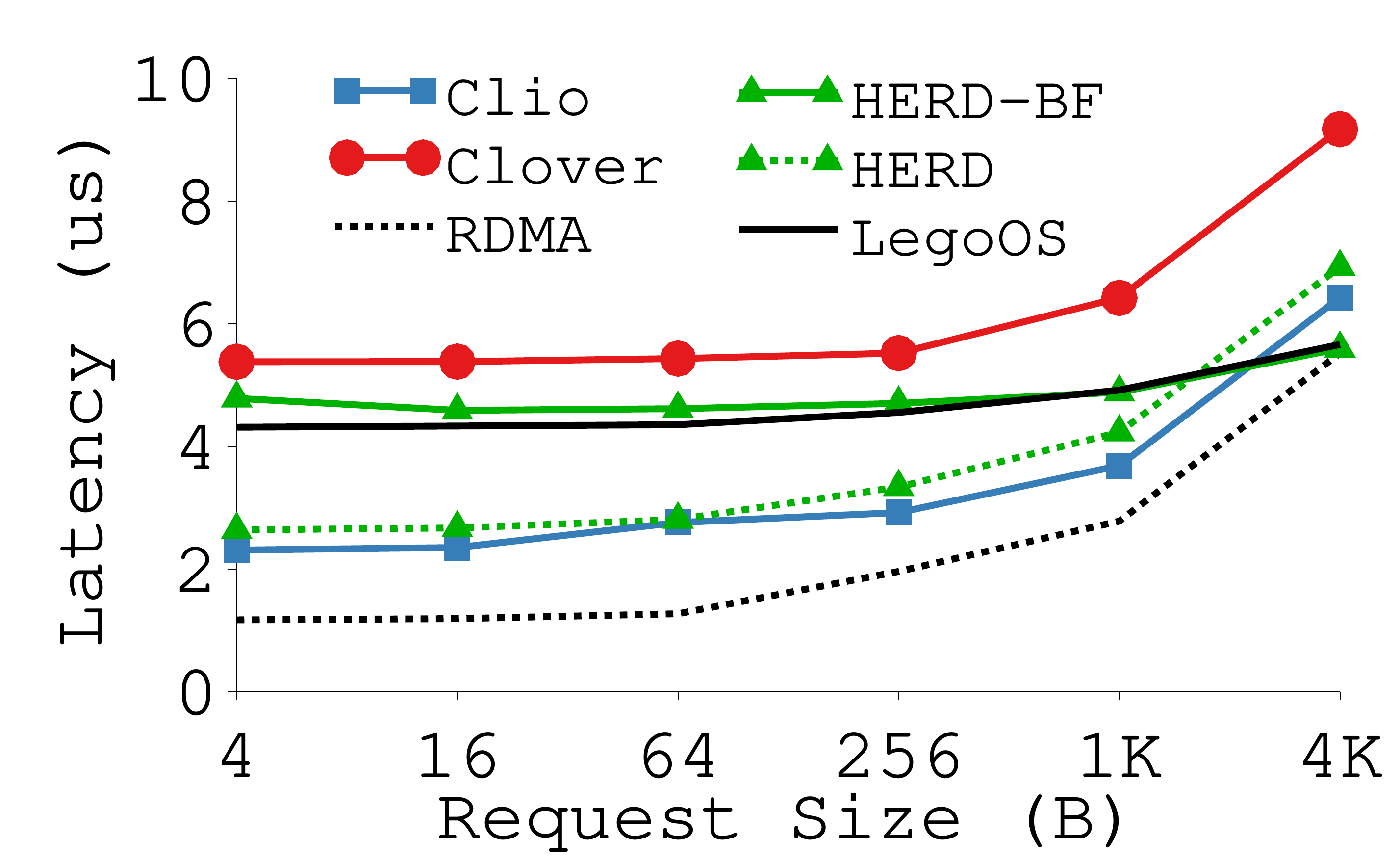}}
\vspace{-0.1in}
\mycaption{fig-write-lat}{Write Latency.}
{
Clover requires $\ge$ 2 RTTs for write.
}
\end{center}
\end{minipage}
\end{figure*}
}

Our evaluation reveals the scalability, throughput, median and tail latency, energy and resource consumption of \sys.
We compare \sys's end-to-end performance with industry-grade NICs (ASIC) and well-tuned RDMA-based software systems.
All \sys's results are FPGA-based, which would be improved with ASIC implementation.

\ulinebfpara{Environment.}
We evaluated \sys\ in our local cluster of four \CN{}s and four \MN{}s (Xilinx ZCU106 boards),
all connected to an Nvidia 40\Gbps\ VPI switch.
Each \CN\ is a Dell PowerEdge R740 server equipped with a Xeon Gold 5128 CPU and a 40\Gbps\ Nvidia ConnectX-3 NIC,
with two of them also having an Nvidia BlueField SmartNIC~\cite{BlueField}.
We also include results from CloudLab~\cite{CloudLab} with the Nvidia ConnectX-5 NIC.

\subsection{Basic Microbenchmark Performance}

\ulinebfpara{Scalability.}
We first compare the scalability of \sys\ and RDMA.
Figure~\ref{fig-conn} measures the latency of \sys\ and RDMA as the number of client processes increases.
For RDMA, each process uses its own QP.
Since \sys\ is connectionless, it scales perfectly with the number of processes.
RDMA scales poorly with its QP, and the problem persists with newer generations of RNIC,
which is also confirmed by our previous works~\cite{Pythia,Storm}.

Figure~\ref{fig-pte-mr} evaluates the scalability with respect to PTEs and memory regions.
For the memory region test, we register multiple MRs using the same physical memory for RDMA.
For \sys, we map a large range of VAs (up to 4\TB) to a small physical memory space, as our testbed only has 2\GB\ physical memory.
However, the number of PTEs and the amount of processing needed are the same for \sysboard\ as if it had a real 4\TB\ physical memory.
Thus, this workload stress tests \sysboard's scalability.
RDMA's performance starts to degrade when there are more than $2^8$ (local cluster) or $2^{12}$ (CloudLab),
and the scalability wrt MR is worse than wrt PTE.
In fact, RDMA fails to run beyond $2^{18}$ MRs.
In contrast, \sys\ scales well and never fails (at least up to 4\TB\ memory).
It has two levels of latency that are both stable: a lower latency below $2^4$ for TLB hit and a higher latency above $2^4$ for TLB miss (which always involves one DRAM access).
A \sysboard\ could use a larger TLB if optimal performance is desired.

These experiments confirm that \textbf{\sys\ can handle thousands of concurrent clients and TBs of memory}.

\ulinebfpara{Latency variation.}
Figure~\ref{fig-miss-hit} plots the latency of reading/writing 16\,B data 
when the operation results in a TLB hit, a TLB miss, a first-access page fault, and MR miss (for RDMA only, when the MR metadata is not in RNIC).
RDMA's performance degrades significantly with misses.
Its page fault handling is extremely slow (16.8\ms).
We confirm the same effect on CloudLab with the newer ConnectX-5 NICs.
\sys\ only incurs a small TLB miss cost and \textbf{no additional cost of page fault handling}.

We also include a projection of \sys's latency if it was to be implemented using a real ASIC-based \sysboard.
Specifically, we collect the latency breakdown of time spent on the network wire and at \CN, time spent on third-party FPGA IPs,
number of cycles on FPGA, and time on accessing on-board DRAM.
We maintain the first two parts, scale the FPGA part to ASIC's frequency (2\,GHz), use DDR access time collected on our server to replace the access time to on-board DRAM (which 
goes through a slow board memory controller).
This estimation is conservative, as a real ASIC implementation of the third-party IPs would make the total latency lower.
Our estimated read latency is better than RDMA, while write latency is worse.
We suspect the reason being Nvidia RNIC's optimization of replying a write before it is fully written to DRAM, which \sys\ could also potentially adopt.

Figure~\ref{fig-tail-latency} plots the request latency CDF of continuously running read/write 16\,B data while not triggering page faults.
Even without page faults, \sys\ has much less latency variation and a much shorter tail than RDMA.

{
\begin{figure*}[th]
\begin{minipage}{\figWidthSix}
\begin{center}
\centerline{\includegraphics[width=\columnwidth]{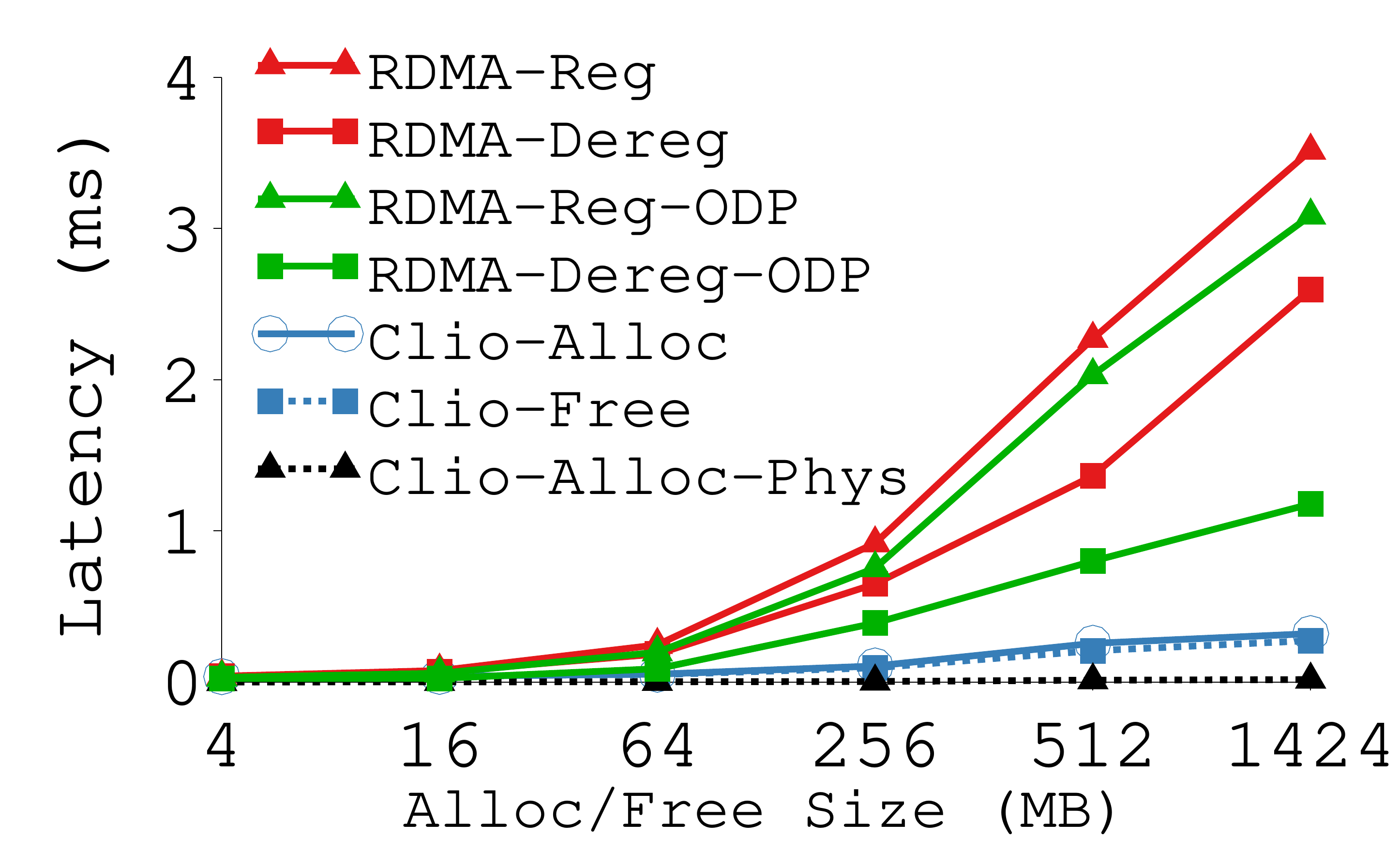}}
\vspace{-0.1in}
\mycaption{fig-alloc-free}{Alloc/Free Latency.}
{
ODP means On-Demand-Paging mode
}
\end{center}
\end{minipage}
\begin{minipage}{\figWidthSix}
\begin{center}
\centerline{\includegraphics[width=\columnwidth]{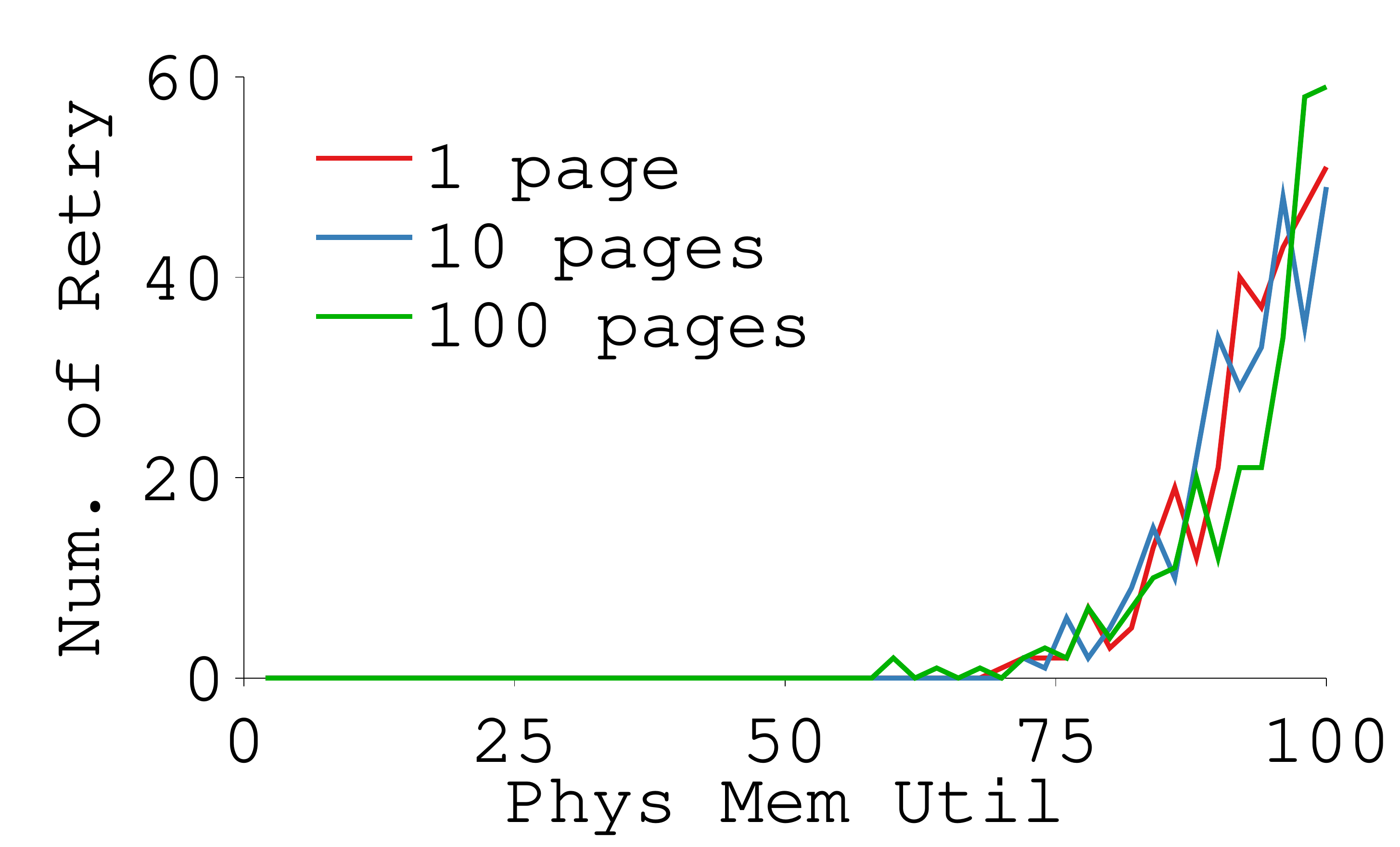}}
\vspace{-0.1in}
\mycaption{fig-alloc-conflict}{Alloc Retry Rate.}
{
}
\end{center}
\end{minipage}
\begin{minipage}{\figWidthSix}
\begin{center}
\centerline{\includegraphics[width=\columnwidth]{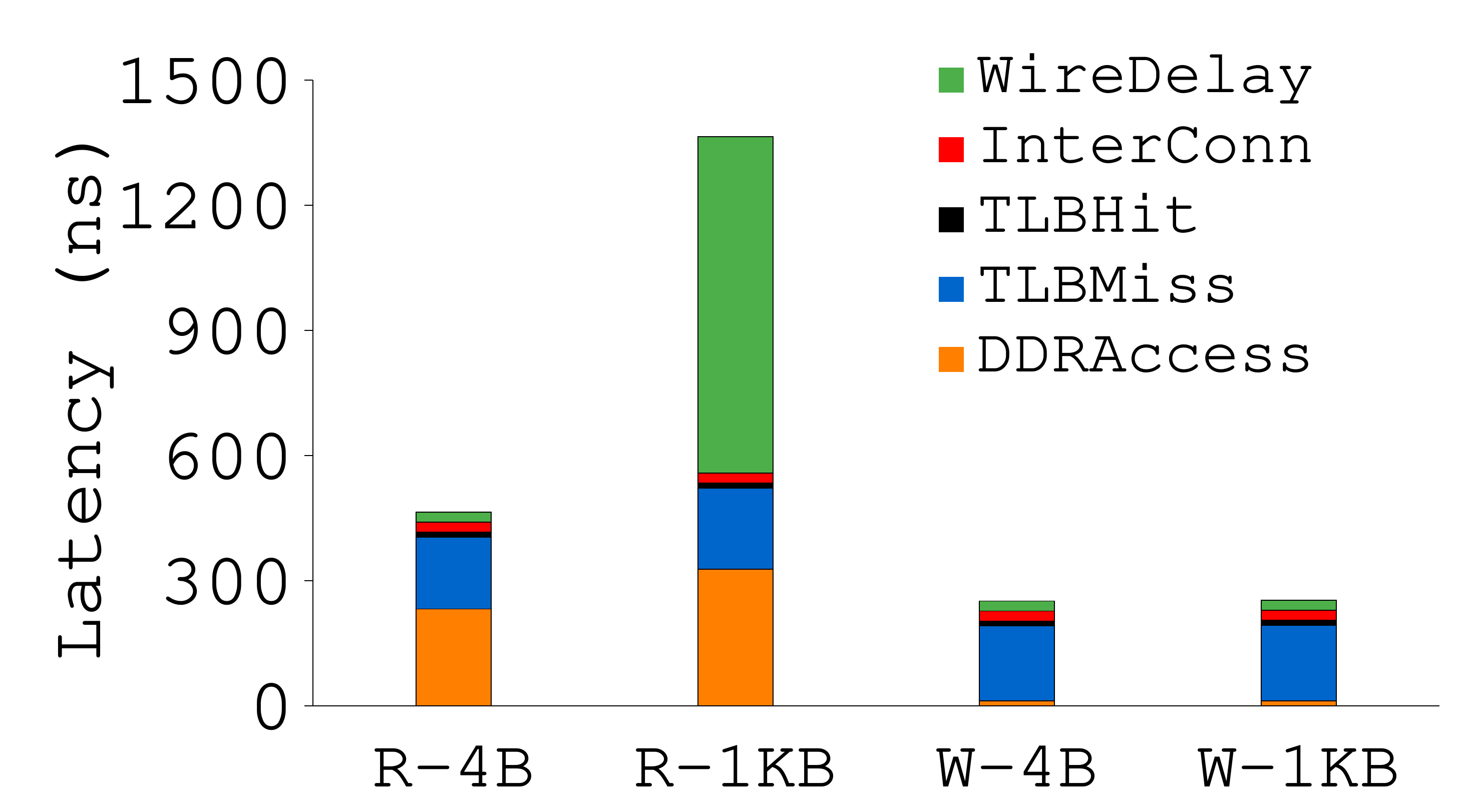}}
\vspace{-0.1in}
\mycaption{fig-lat-break}{Latency Breakdown.}
{
Breakdown of time spent at \sysboard.
}
\end{center}
\end{minipage}
\begin{minipage}{\figWidthSix}
\begin{center}
\centerline{\includegraphics[width=\columnwidth]{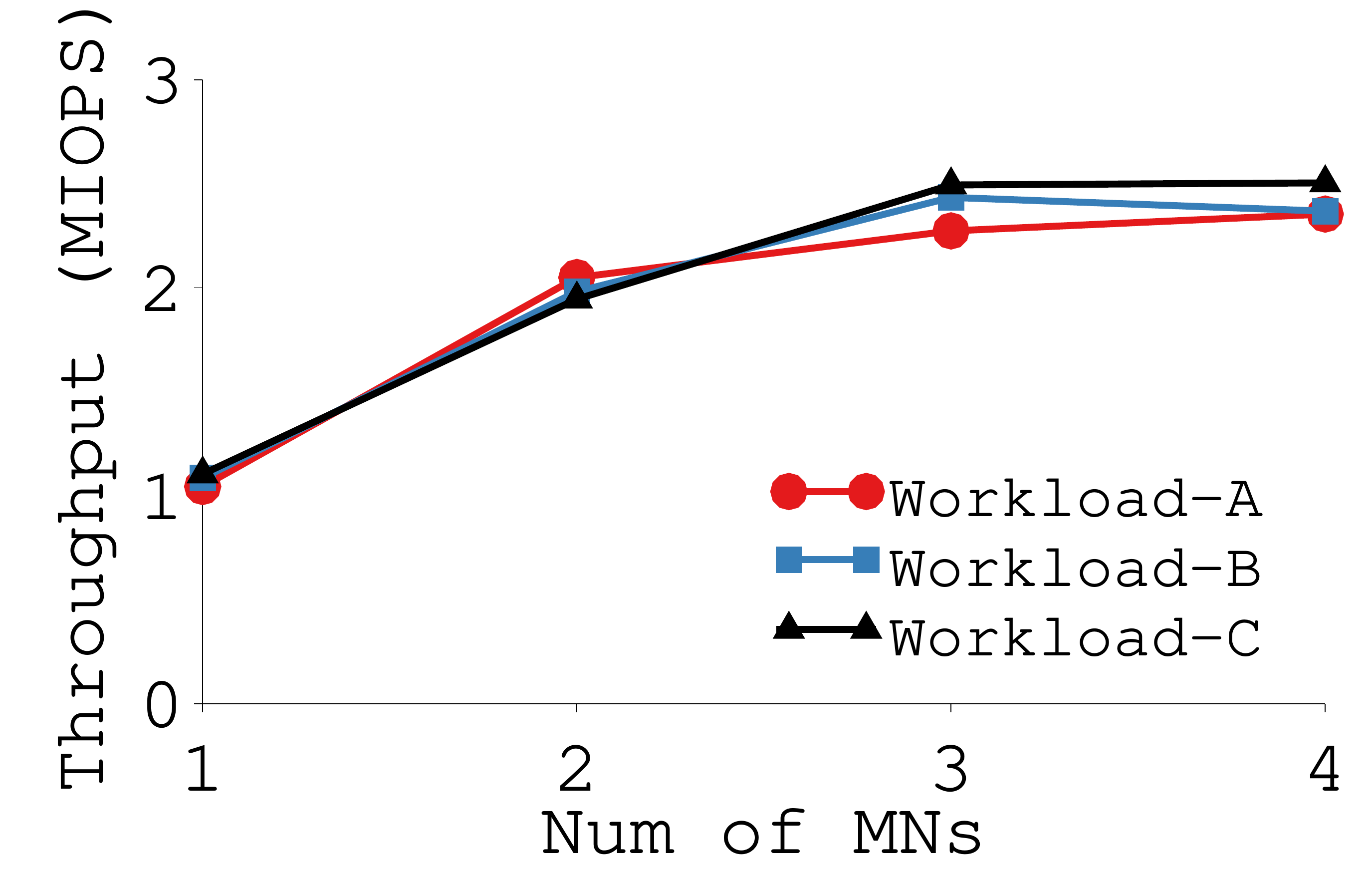}}
\vspace{-0.1in}
\mycaption{fig-ycsb-mn}{\syskv\ Scalability against \MN{}s.}
{
}
\end{center}
\end{minipage}
\vspace{-0.15in}
\end{figure*}
}

{
\begin{figure*}[th]
\begin{minipage}{\figWidthSix}
\begin{center}
\centerline{\includegraphics[width=\columnwidth]{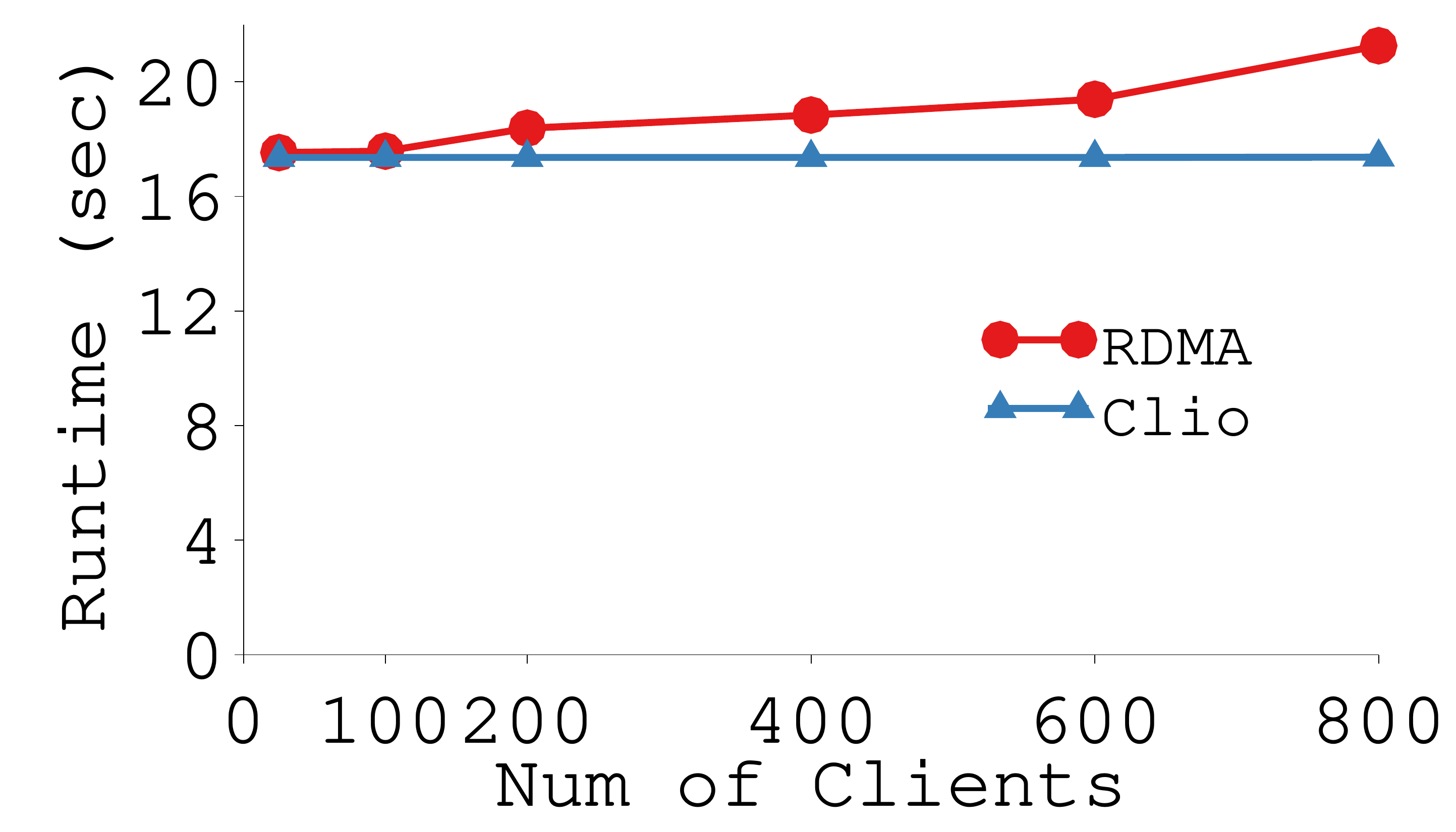}}
\vspace{-0.1in}
\mycaption{fig-photo}{Image Compression.}
{
}
\end{center}
\end{minipage}
\begin{minipage}{\figWidthSix}
\begin{center}
\centerline{\includegraphics[width=\columnwidth]{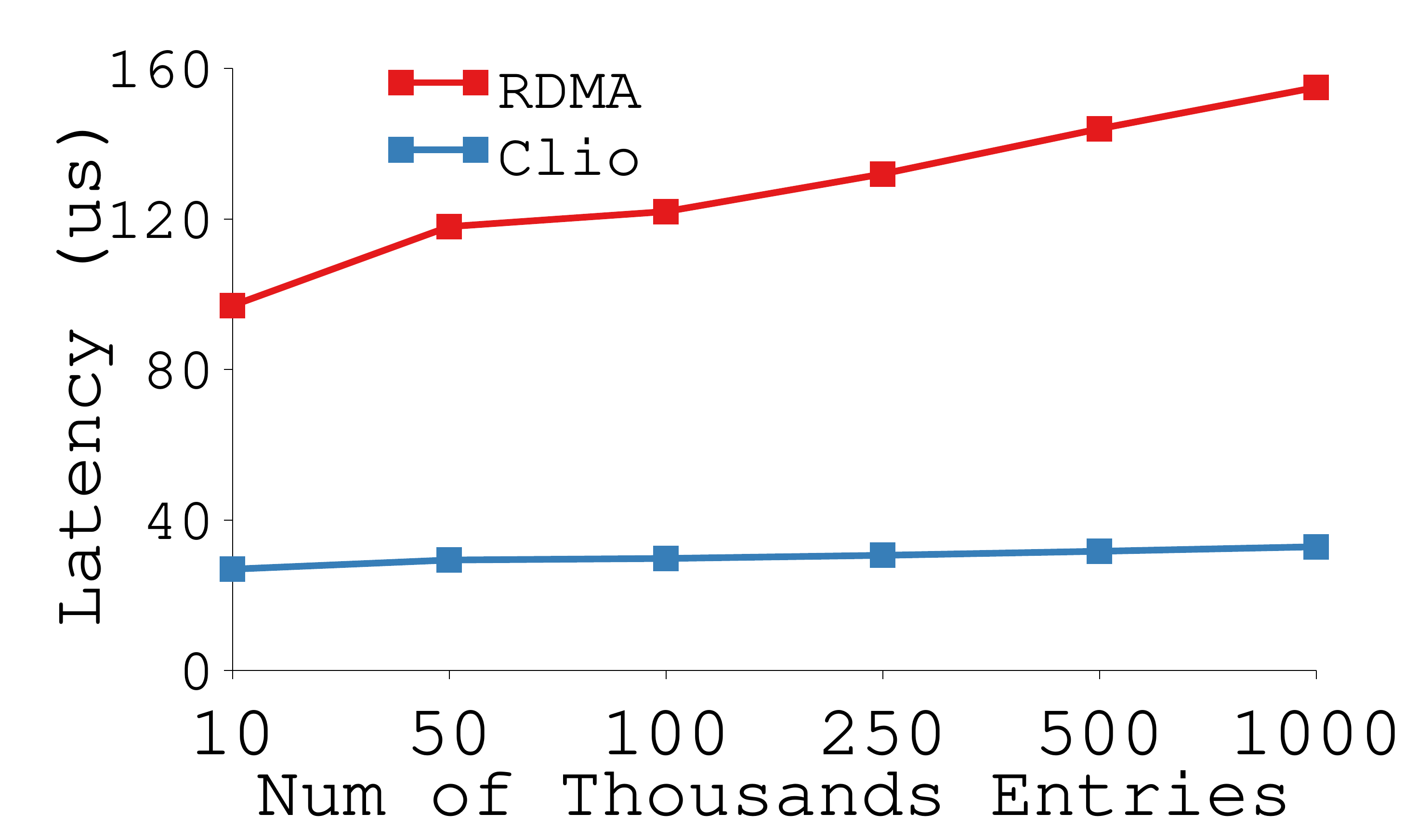}}
\vspace{-0.1in}
\mycaption{fig-radix}{Radix Tree Search Latency.}
{
}
\end{center}
\end{minipage}
\vspace{-0.1in}
\begin{minipage}{\figWidthSix}
\begin{center}
\centerline{\includegraphics[width=\columnwidth]{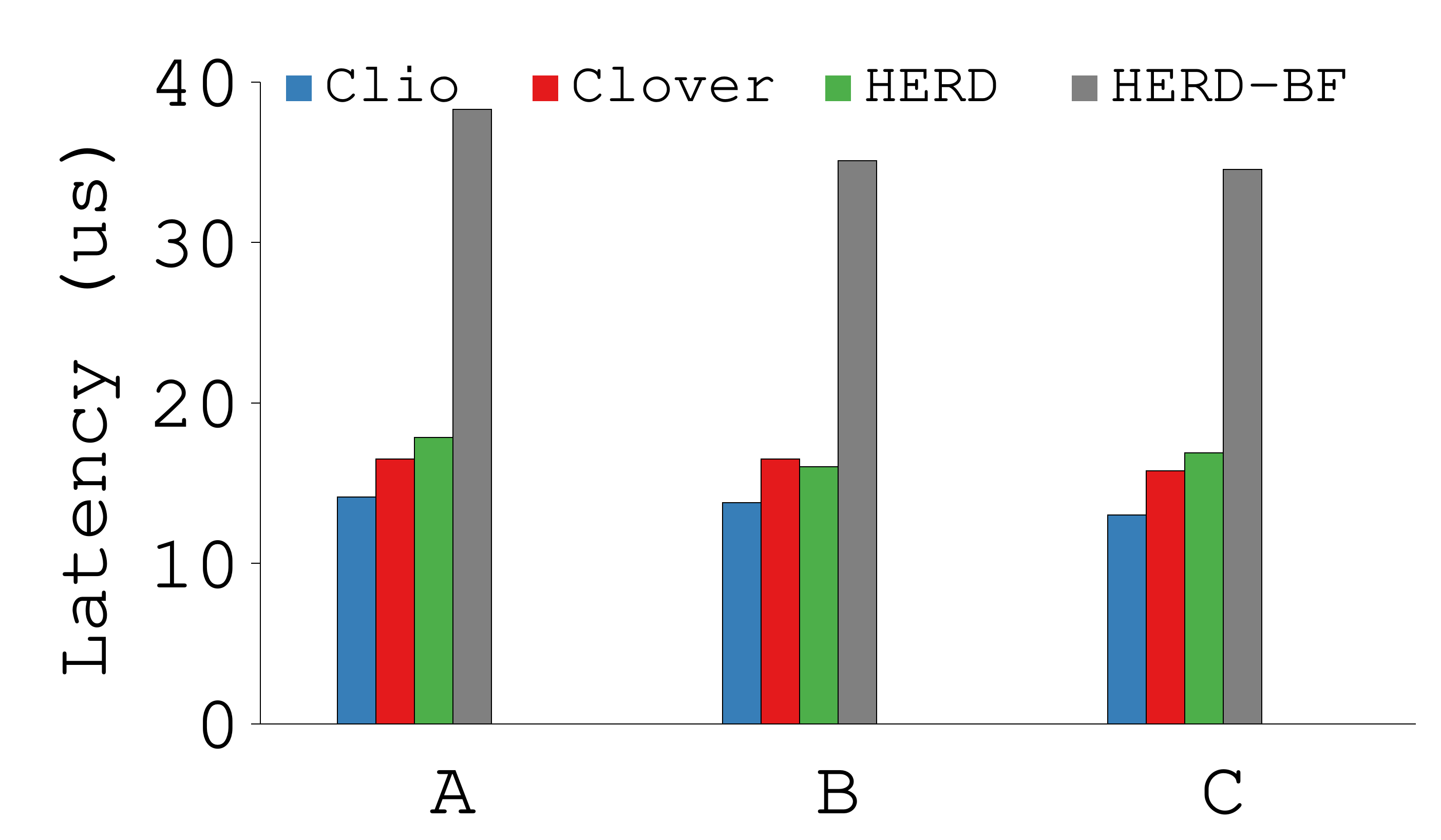}}
\vspace{-0.1in}
\mycaption{fig-kvstore}{Key-Value Store YCSB Latency.}
{
}
\end{center}
\end{minipage}
\vspace{-0.1in}
\begin{minipage}{\figWidthSix}
\begin{center}
\centerline{\includegraphics[width=\columnwidth]{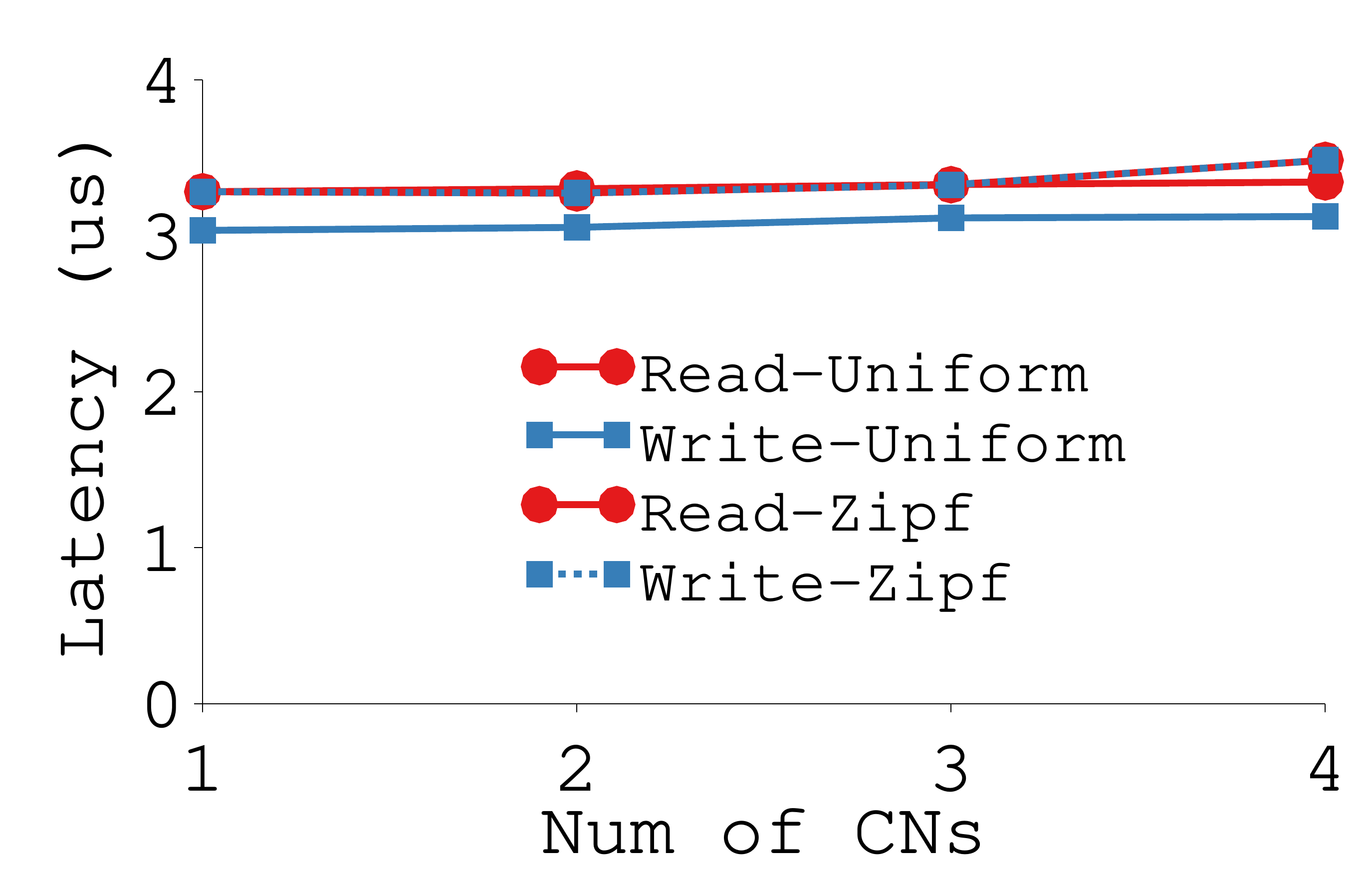}}
\vspace{-0.1in}
\mycaption{fig-mvstore}{\sysmv\ Object Read/Write Latency.}
{
}
\end{center}
\end{minipage}
\end{figure*}
}

\ulinebfpara{Read/write throughput.}
We measure \sys's throughput by varying the number of concurrent client threads (Figure~\ref{fig-read-write-throughput}).
\sys's default asynchronous APIs quickly reach the line rate of our testbed (9.4\Gbps\ maximum throughput).
Its synchronous APIs could also reach line rate fairly quickly.

Figure~\ref{fig-onboard-throughput} measures the maximum throughput of \sys's FPGA implementation without the bottleneck of the board's 10\Gbps\ port, by generating traffic on board.
Both read and write can reach more than 110\Gbps\ when request size is large.
Read throughput is lower than write when request size is smaller.
We found the throughput bottleneck to be the third-party non-pipelined DMA IP
(which could potentially be improved).

\ulinebfpara{Comparison with other systems.}
We compare \sys\ with native one-sided RDMA, Clover~\cite{Tsai20-ATC}, HERD~\cite{Kalia14-RDMAKV}, and LegoOS~\cite{Shan18-OSDI}.
We ran HERD on both CPU and BlueField (HERD-BF).
Clover is a passive disaggregated persistent memory system which we adapted as a passive disaggregated memory (PDM) system.
HERD is an RDMA-based system that supports a key-value interface with an RPC-like architecture.
LegoOS builds its virtual memory system in software at \MN.

\sys's performance is similar to HERD and close to native RDMA.
Clover's write is the worst because it uses at least 2 RTTs for writes to deliver its consistency guarantees without any processing power at \MN{}s.
HERD-BF's latency is much higher than when HERD runs on CPU
due to the slow communication between BlueField's ConnectX-5 chip and ARM processor chip.
LegoOS's latency is almost two times higher than \sys's when request size is small.
In addition, from our experiment, LegoOS can only reach a peak throughput of 77\Gbps, while \sys\ can reach 110\Gbps.
LegoOS' performance overhead comes from its software approach, demonstrating the necessity of a hardware-based solution like \sys.

\ulinebfpara{Allocation performance.}
Figure~\ref{fig-alloc-free} shows \sys's VA and PA allocation and RDMA's MR registration performance.
\sys's PA allocation takes less than 20\mus, and the VA allocation is much faster than RDMA MR registration,
although both get slower with larger allocation/registration size.
Figure~\ref{fig-alloc-conflict} shows the number of retries at allocation time with three allocation sizes as the physical memory fills up.
There is no retry when memory is below half utilized. Even when memory is close to full, there are at most 60 retries per allocation request, with roughly 0.5\ms\ per retry. This confirms that our design of avoiding hash overflows at allocation time is practical.

\ulinebfpara{Close look at \sysboard{} components.}
To further understand \sys's performance, 
we profile different parts of \sys's processing for read and write of 4\,B to 1\KB.
\syslib\ adds a very small overhead (250\ns\ in total), 
thanks to our efficient threading model and network stack implementation.
Figure~\ref{fig-lat-break} shows the latency breakdown at \sysboard.
Time to fetch data from DRAM (DDRAccess) and to transfer it over the wire (WireDelay) are the main 
contributor to read latency, especially with large read size.
Both could be largely improved in a real \sysboard\ with better memory controller and higher frequency.
TLB miss (which takes one DRAM read) is the other main part of the latencies.

\subsection{Application Performance}

\ulinebfpara{Image Compression.}
We run a workload where each client 
compresses and decompresses 1000 256*256-pixel images with increasing number of concurrently running clients.
Figure~\ref{fig-photo} shows the total runtime per client.
We compare \sys\ with RDMA, with both performing computation at the \CN\ side and the RDMA using one-sided operations instead of \sys\ APIs to read/write images in remote memory.
\sys's performance stays the same as the number of clients increase.
RDMA's performance does not scale because it requires each client to register a different MR to have protected memory accesses.
With more MRs, RDMA runs into the case where the RNIC cannot hold all the MR metadata and many accesses would involve a slow read to host main memory.

\ulinebfpara{Radix Tree.}
Figure~\ref{fig-radix} shows the latency of searching a key in pre-populated radix trees when varying the tree size. 
We again compare with RDMA which uses one-sided read operations to perform the tree traversal task.
RDMA's performance is worse than \sys,
because it requires multiple RTTs to traverse the tree,
while \sys\ only needs one RTT for each pointer chasing (each tree level).
In addition, RDMA also scales worse than \sys.

\ulinebfpara{Key-value store.}
Figure~\ref{fig-kvstore} evaluates \syskv\ using the YCSB benchmark~\cite{YCSB} and compares it to Clover, HERD, and HERD-BF.
We run two \CN{}s and 8 threads per \CN.
We use 100K key-value entries and run 100K operations per test,
with YCSB's default key-value size of 1\KB. 
The accesses to keys follow the Zipf distribution ($\theta=0.99$).
We use three YCSB workloads with different {\em get-set} ratios: 
100\% {\em get} (workload C), 5\% {\em set} (B), and 50\% {\em set} (A).
\syskv\ performs the best.
HERD running on BlueField performs the worst, mainly because BlueField's slower crossing between its NIC chip and ARM chip.

{
\begin{figure*}[th]
\begin{minipage}{\figWidthSix}
\begin{center}
\centerline{\includegraphics[width=\columnwidth]{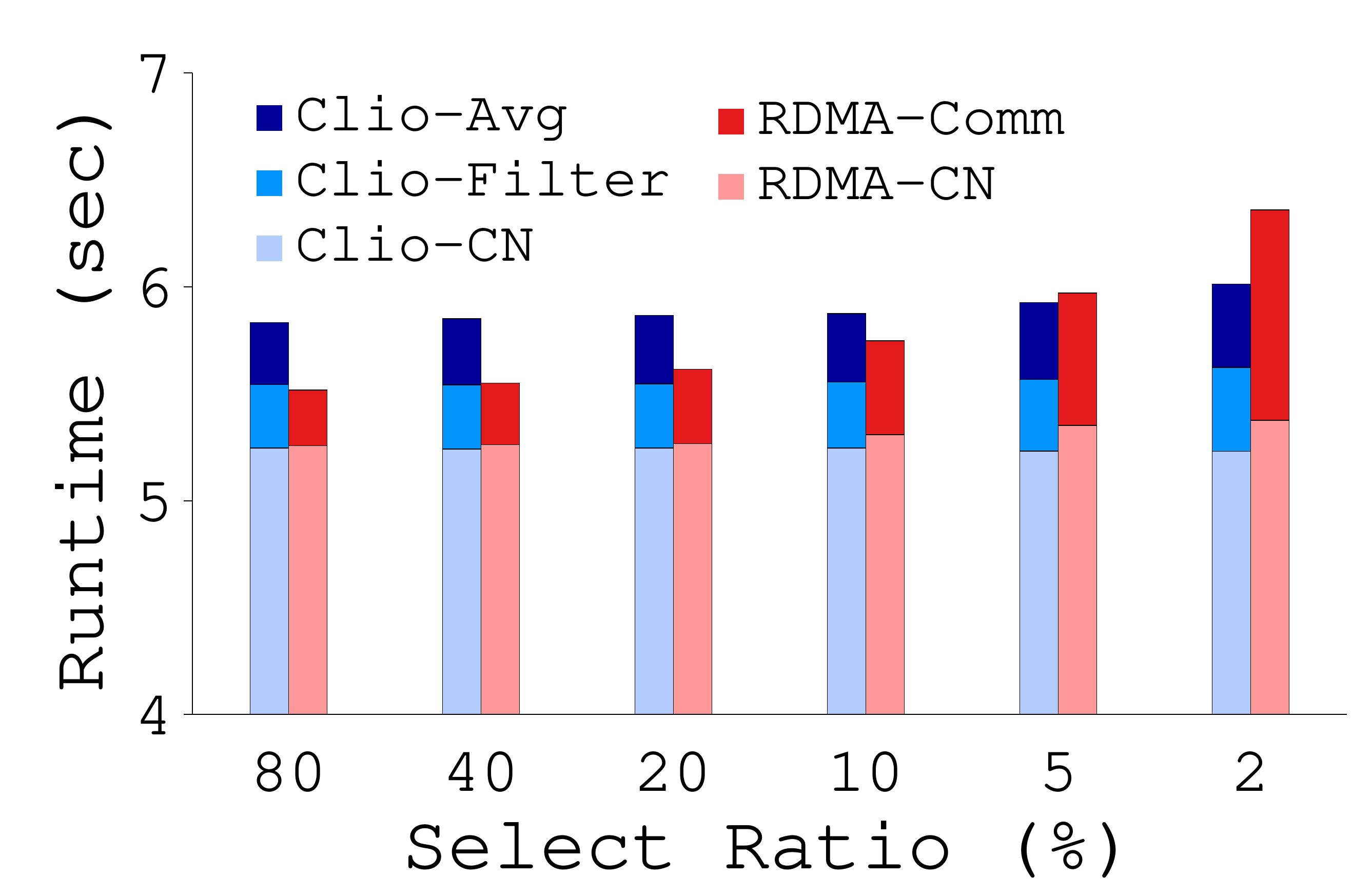}}
\vspace{-0.1in}
\mycaption{fig-dataframe}{Select-Aggregate-Shuffle.}
{
Y axis starts at 4 sec. 
CN represents computation done at \CN.
}
\end{center}
\end{minipage}
\begin{minipage}{0.2in}
\hspace{0.2in}
\end{minipage}
\begin{minipage}{\figWidthSix}
\begin{center}
\centerline{\includegraphics[width=\columnwidth]{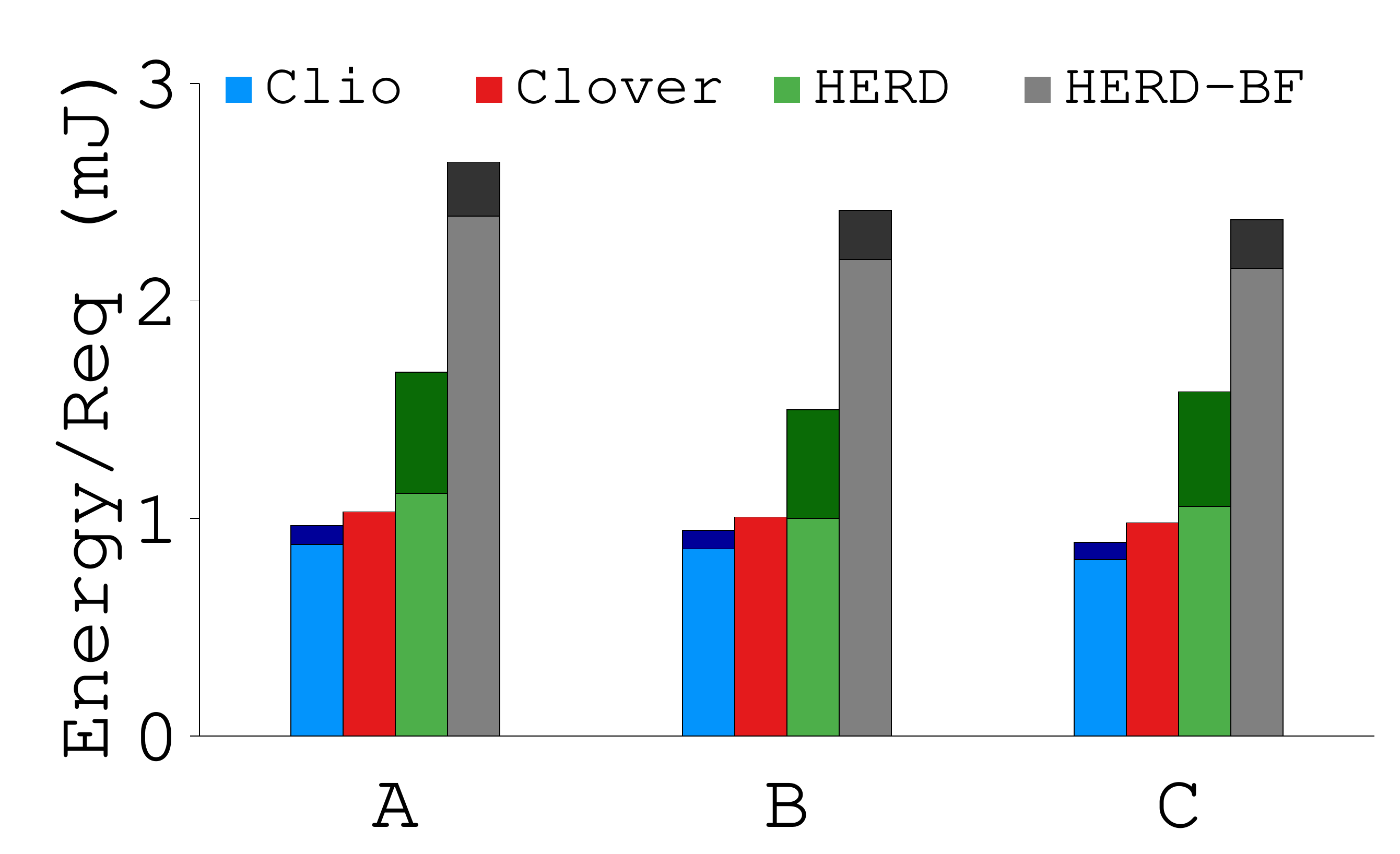}}
\vspace{-0.1in}
\mycaption{fig-energy}{Energy Comparison.}
{
Darker/lighter shades represent energy spent at \MN{}s and \CN{}s.
}
\end{center}
\end{minipage}
\begin{minipage}{0.2in}
\hspace{0.2in}
\end{minipage}
\begin{minipage}{\figWidthSix}
\begin{center}
\scriptsize
\begin{tabular}{ p{0.7in} | p{0.2in} |p{0.27in} }

 & \textbf{Logic} & \textbf{Memory} \\
\textbf{System/Module} & \textbf{(LUT)} & \textbf{(BRAM)} \\
\hline
\hline
StRoM-RoCEv2 & 39\% & 76\% \\
Tonic-SACK & 48\% & 40\% \\
\hline
\sys\ (Total) & 31\% & 31\% \\
VirtMem & 5.5\% & 3\% \\
NetStack & 2.3\% & 1.7\% \\
\hline
Go-Back-N & 5.8\% & 2.6\% \\

\end{tabular}
\mycaption{fig-fpga-resource}{FPGA Utilization.}
{
}
\end{center}
\end{minipage}
\end{figure*}
}

Figures~\ref{fig-ycsb-mn} shows the throughput of \syskv\ when varying the number of MNs. Similar to our
\sys\ scalability results, \syskv\ can reach a CN’s maximum
throughput and can handle concurrent get/set requests even
under contention. These results are similar to or better than
previous FPGA-based and RDMA-based key-value stores that
are fine-tuned for just key-value workloads (Table 3 in \cite{KVDIRECT}),
while we got our results without any performance tuning.

\ulinebfpara{Multi-version data store.}
We evaluate \sysmv\ by varying the number of \CN{}s that concurrently access data objects (of 16\,B) on an \MN\ using workloads of 50\% read (of different versions) and 50\% write under uniform and Zipf distribution of objects (Figure~\ref{fig-mvstore}). 
\sysmv's read and write have the same performance, and reading any version has the 
same performance, since we use an array-based version design. 

\ulinebfpara{Data analytics.}
We run a simple workload which first \texttt{select} rows in a table whose field-A matches a value (\eg, gender is female)
and calculate \texttt{avg} of field-B (\eg, final score) of all the rows.
Finally, it calculates the histogram of the selected rows (\eg, score distribution), which can be presented to the user together with the avg value (\eg, how female students' scores compare to the whole class).
\sys\ executes the first two steps at \MN\ offloads and the final step at \CN,
while RDMA always reads rows to \CN\ and then does each operation.
Figure~\ref{fig-dataframe} plots the total run time as the select ratio decreases (fewer rows selected).
When the select ratio is high, \sys\ and RDMA send a similar amount of data across the network,
and as the CPU computation is faster than our FPGA implementation for these operations, \sys's overall performance is worse than RDMA.
When the select ratio is low, \sys\ transfers much less data than RDMA, resulting in its better performance.

\subsection{CapEx, Energy, and FPGA Utilization}
\label{sec:results-cost}



We estimate the cost of server and \sysboard\ using market prices of different hardware units. When using 1\TB\ DRAM, a server-based \MN\ costs 1.1-1.5\x\ and consumes 1.9-2.7\x\ power compared to \sysboard. These numbers become 1.4-2.5\x\ and 5.1-8.6\x\ with OptaneDimm~\cite{optane-dcpm}, which we expect to be the more likely remote memory media in future systems.

We measure the total energy used for running YCSB workloads
by collecting the total CPU (or FPGA) cycles and the Watt of a CPU core~\cite{gold5128}, ARM processor~\cite{armpower}, and FPGA (measured).
We omit the energy used by DRAM and NICs in all the calculations. 
Clover, a system that centers its design around low cost, has slightly higher energy than \sys.
Even though there is no processing at \MN{}s for Clover, its \CN{}s use more cycles to process and manage memory.
HERD consumes 1.6\x\ to 3\x\ more energy than \sys, mainly because of its CPU overhead at \MN{}s.
Surprisingly, HERD-BF consumes the most energy, even though it is a low-power ARM-based SmartNIC.
This is because of its worse performance and longer total runtime.

Figure~\ref{fig-fpga-resource} compares the FPGA utilization among Clio, StRoM's RoCEv2~\cite{StRoM}, and Tonic's selective ack stack~\cite{TONIC}.
Both StRoM and Tonic include only a network stack but they consume more resources than \sys.
Within \sys, the virtual memory (VirtMem) and
the network stack (NetStack) consume a small fraction of the total resources,
with the rest being vendor IPs (PHY, MAC, DDR4, and interconnect).
Overall, our efficient hardware implementation leaves most FPGA resources available for application offloads.

\section{Discussion and Conclusion}
\label{sec:discussion}

We presented \sys, a new hardware-based disaggregated memory system.
Our FPGA prototype demonstrates that \sys\ achieves great performance, scalability, and cost-saving.
This work not only guides the future development of \md\ solutions
but also demonstrates how to implement a core OS subsystem in hardware and co-design it with the network.
We now present our concluding thoughts.


\ulinebfpara{Security and performance isolation.}
\sys’s protection domain is a user process, which is the same as the traditional single-server process-address-space-based protection. The difference is that \sys\ performs permission checks at MNs: it restricts a process’ access to only its (remote) memory address space and does this check based on the global PID. Thus, the safety of \sys\ relies on PIDs to be authentic (\eg, by letting a trusted CN OS or trusted CN hardware attach process IDs to each \sys\ request). There have been researches on attacking RDMA systems by forging requests~\cite{ReDMArk-security21} and on adding security features to RDMA~\cite{1RMA,sRDMA-ATC20}. How these and other existing security works relate and could be extended in a memory disaggregation setting is an open problem, and we leave this for future work.

There are also designs in our current implementation that could be improved to provide more protection against side-channel and DoS attacks.
For example, currently, the TLB is shared across application processes,
and there is no network bandwidth limit for an individual connection.
Adding more isolation to these components would potentially increase the cost of \sysboard\ or reduce its performance.
We leave exploring such tradeoffs to future work.

\ulinebfpara{Failure handling.}
Although memory systems are usually assumed to be volatile, 
there are still situations that require proper failure handling (\eg, for high availability or to use memory for storing data).
As there can be many ways to build memory services on \sys\ 
and many such services are already or would benefit from handling failure on their own,
we choose not to have any built-in failure handling mechanism in \sys.
Instead, \sys\ should offer primitives like replicated writes for users to build their own services.
We leave adding such API extensions to \sys\ as future work.

\ulinebfpara{\CN-side stack.}
An interesting finding we have is that \CN-side systems
could become a performance bottleneck after we made the remote memory layer very fast.
Surprisingly, most of our performance tuning efforts are spent on the \CN\ side (\eg, thread model, network stack implementation).
Nonetheless, software implementation is inevitably slower than customized hardware implementation.
Future works could potentially improve \sys's \CN\ side performance by offloading the software stack to a customized hardware NIC.

\section*{Acknowledgement}

We would like to thank the anonymous reviewers and our shepherd Mark Silberstein
for their tremendous feedback and comments, which have
substantially improved the content and presentation of this paper.
We are also thankful to Geoff Voelker, Harry Xu, Steven Swanson, Alex Forencich for their valuable feedback on our work.

This material is based upon work supported by the National
Science Foundation under the following grant: NSF 2022675, and gifts from VMware.
Any opinions, findings, and conclusions or recommendations
expressed in this material are those of the authors and do not 
necessarily reflect the views of NSF or other institutions.

\balance
\bibliographystyle{ACM-Reference-Format}
\bibliography{all-defs,all,personal,all-confs,local,paper}

\end{document}


\begin{appendices}

\section{\phdm\ Use Cases}
Many types of applications can make use of \phdm.
Below we give some examples. 
We implemented an instance of the first three types in this paper,
leaving the rest for future work.

\boldpara{Extended (semantic-rich) virtual memory.}
A basic service \phdm\ can provide is a remote virtual memory space that lets applications
store in-memory data (\eg, as extended, slower heaps).
In addition to simple, hardware-like virtual memory APIs such as reading and writing to a memory address, 
\phdm\ could provide higher-level APIs like synchronization primitives, pointer manipulation, 
vector and scatter-gather operations~\cite{Aguilera-FarMemory}.
Applications and language libraries can then build complex data structures like vectors 
and trees with these APIs.

\boldpara{In-memory and ephemeral storage.}
\phdm\ could offer in-memory storage services such as distributed key-value stores, databases, and file systems.
With \phdm, many storage operations (\eg, key-value pair lookup, SQL select) 
could be implemented in hardware at where the data is, offering enhanced performance. 
\phdm\ is also a good fit for building ephemeral storage and storage caching that do not require failure resilience~\cite{SnowFlake-NSDI20,Pocket,fitzpatrick2004distributed}.

\boldpara{Data sharing.}
Since multiple \CN{}s can access the same \MN{}s in \phdm,
\phdm\ could be used for data sharing and communication across \CN{}s.
This is especially useful for new datacenter services like serverless computing~\cite{Berkeley-Serverless},
which currently has no or poor support for managing states and inter-function communication.
With \phdm, serverless functions can run on \CN{}s and store states or communication messages in the disaggregated memory layer.
Similarly, \phdm\ can also be used for storing global states such as the parameter server in distributed machine learning systems.

\boldpara{Offloading data processing.}
\phdm\ is a good candidate for offloading data processing and data analytics. 
Data-intensive applications can offload computation that frequently access in-memory data together with 
these data to \MN{}s.
One such example is disaggregated Spark shuffle~\cite{Stuedi-ATC19}, where the shuffle
operation could be implemented in programmable hardware and the shuffle data could be 
stored in \MN{}s of \phdm.

\boldpara{Remote swap and remote disk.}
Legacy applications and libraries can also benefit from \phdm\ in a transparent way.
Two such examples are remote memory-based disk and remote swap~\cite{InfiniSwap}.
The OS at the \CN{}s can add a memory-based block device that sits in \phdm\
in a similar way as building the {\em ramdisk} module.
Applications can directly use this new device or use it as a swap space.

\boldpara{Disaggregated OS.}
Recently, there have been proposals to completely disaggregate memory from compute.
\lego~\cite{Shan18-OSDI} is such a proposal that organizes compute nodes as processors with no memory 
and \MN{}s as memory devices with no computation.
\lego\ can build on \phdm\ by configuring \CN{}s as its compute nodes 
and \MN{}s as its memory nodes. 

\section{Discussion}
\label{sec:discussion}

\ulinebfpara{Hardware choice.}
A key contribution of \sys\ is our demonstration of how to best separate 
data and metadata planes into hardware and software components.
This conceptual separation could be adopted by future system builders
in addition to our specific implementation of \sys.
Although we developed \sys's hardware components all on a single FPGA,
we do not see it as the only way to build the hardware part of \sys.
Another promising approach is to separate out the fixed logics in \sys\ hardware components
into an ASIC and the rest in FPGA. 
For example, for environments that do not need customization of network or the core-memory stack,
they could be integrated into one ASIC based network controller to provide fast network/memory accesses,
ASIC based memory controller IPs can also be added to provide modern CPU level memory access performance.
while extended services are still deployed on an FPGA.
Note that this ASIC would still be very different from today's RDMA NIC,
as our stacks have a client-centric addressing view.
Also note that there are still cases where users would want to customize \sys's network and the core-memory stack,
\eg, to provide better performance isolation and security.


\ulinebfpara{Security and performance isolation.}
\sys\ currently adopts traditional address-space-based protection mechanisms to 
isolate application processes from each other 
and to restrict non-\sys\ owners from changing \sys\ or other user's hardware and software.
When deployed in the cloud, multi-tenancy environment that requires stronger security and/or performance isolation (\eg, to prevent side-channel attacks),
and our current implementation of \sys\ needs some adaptation. 
For example, the PTE cache currently is shared across application processes,
and there is no network bandwidth limits for an individual connection.
It is fairly easy to change them to provide more isolation and SLA guarantees,
and we leave it for future work. 

\ulinebfpara{Failure handling.}
Although memory systems are usually assumed to be volatile, 
there are still situations that require proper failure handling (\eg, for high availability or to use memory for storage data).
As there can be many ways to build memory services on \sys\ 
and many such services are already or would benefit from handling failure on their own,
we believe that there should not be any built-in failure handling mechanism in \sys.
Instead, \sys\ should offer primitives like replicated writes for users to build their own services.
We leave adding such API extensions to \sys\ as future work.

\ulinebfpara{Client-side.}
An interesting finding we have is that client-side systems
could become a performance bottleneck after we made the remote memory layer very fast.
Surprisingly, most of our performance tuning efforts are spent on the client side (\eg, thread model, network stack implementation).
More future research could go into finding the best way for client applications to fully exploit \sys's remote memory performance.

\section{\sys\ Address Translation}

Figure~\ref{fig-coremem-diagram} shows the detailed state diagram of \sys's FPGA
core-memory module.

{
\begin{figure}[!ht]
\begin{minipage}{\columnwidth}
\begin{center}
\centerline{\includegraphics[width=\textwidth]{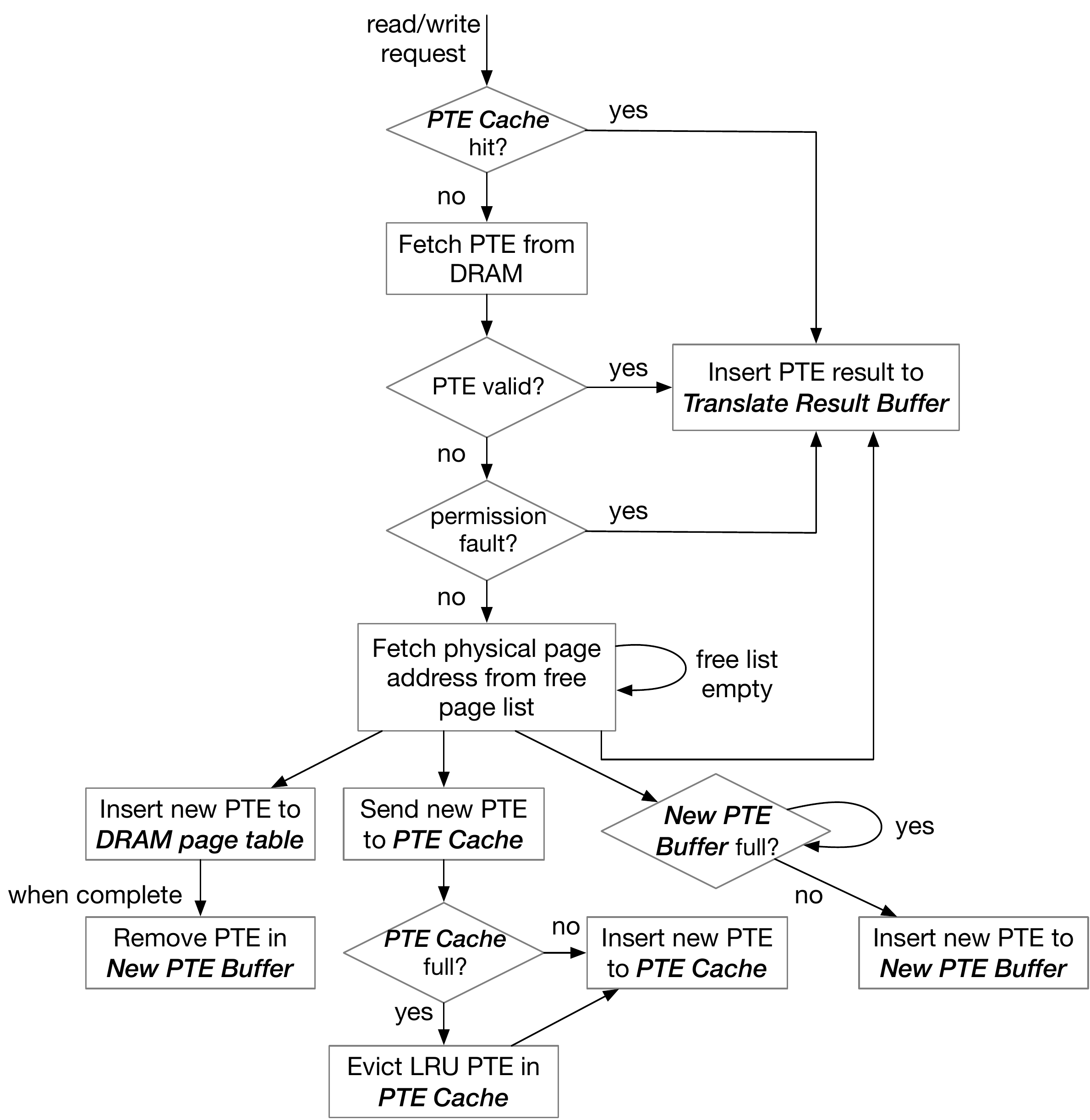}}
\mycaption{fig-coremem-diagram}{Core-Memory Address Translation State Diagram.}
{
}
\end{center}
\end{minipage}
\end{figure}
}

\end{appendices}

\bibliographystyle{plain}
\bibliography{all-defs,all,personal,all-confs,local,paper}